\documentclass[aps,prd]{revtex4}
\newcommand{\be}{\begin{equation}}
\newcommand{\ee}{\end{equation}}
\newcommand{\ba}{\begin{eqnarray}}
\newcommand{\ea}{\end{eqnarray}}

\begin{document}

\title{COMMENTS ON SPACE-TIME}

\author{K. Ghosh\footnote{E-address: kaugho@rediffmail.com}} 
\affiliation{A.L - 123, Sector -2, Bidhan nagar, Kolkata-700 091, India.}

\maketitle

\section{Abstract}

In this article we discuss a few aspects of the space-time
description of fields and particles. In sectionn II and III
we demonstrate that fields are as fundamental as particles. 
In section IV we discuss non-equivalence
of the Schwarzschild coordinates and the Kruskal-Szekeres
coordinates. In section V we discuss that it is not
possible to define causal structure in discrete space-time
manifolds. In App.B we show that a line is not just a collection
of points and we will have to introduce one-dimensional
line-intervals as fundamental geometric elements. Similar
discussions are valid for area and volume-elements.
In App.C and App.D we make a comparative study of Quantum
Field Theory and Quantum Mechanics and contradictions
associated with probabilistics interpretation of these
theories with space-time dimensional analysis. In
App.E and App.F we discuss the geometry of Robertson-Walker model
and electrostatic behavior of dielectrics respectively.
In Sup.I we discuss the regularity of Spin-Spherical
harmonics and also derive an energy-spectrum which
is free of back-reaction problem. In Sup.II we discuss that in general the integral version
of Gauss's divergence law in Electrodynamics is not valid
and rederive Gauss's law and Ampere's law. We also show
that under duality transformation magnetic charge
conservation law do not remain time reversal symmetric.
In Sup.III we derive the complete equation for viscous
compressible fluids and make a few comments regarding
some cotradictions associated with boundary conditions
for fluid dynamics. 
In Sup.IV we discuss a few aspects on double slit
interference experiments.
We conclude this article with a few questions in
Sup.V.


\section{{Overview}}

Gravitational interaction which is universally attractive
is described by the General Theory of Relativity, a theory
of fields realized through the existence of space-time and
some of its geometric properties,e.g, the curvature. The principle
of equivalence leading to the fact that the guinea and the
feather fall the same way in vacuum together with the
fact that gravity violets the elementary quantum
mechanical priciples (as will be evident from this article)
indicates that fields are as fundamental as particles. Fields
and particles with their common and contradictory kinemetical
and dynamical features are the two fundamental constituents
of nature. The attraction of opposite charges can only be
explained through accepting the electric field as a fundamental
entity of nature and not through only 
interaction mediating particle interaction
(be the interaction mediating particles interacting
with the sources or among themselves)
although
the repulsion between like charges may be explained in
the later way with photons as the interaction mediator.
Quantum mechanics is an incomplete approach to explain
thr Solar system microscopic physics purely in terms of wave-like
properties whereas Quantum Field Theory is an incomplete
approach to explain
the Solar system microscopic physics purely in terms of 
particle-like properties.
For the vacuum polarization explanation
of the potential [18] in the Bhaba scattering process
it is not obvious how, in terms of photon
exchanges, the loops in vacuum
will screen the charges of $e^- ,e^+$.
Also an electron-positron loop with two
external photon lines can not explain
charge screening in any process as the
particle-antiparticle pairs are created and
annihilated after the interaction mediating photon had been created
at one real particle and before the interaction mediating photon has
interacted with the other real particle.
None of these two theories
are in accordance with gravity as will be manifested
in this article.

\section{{Introduction}}

During the last few decades a lot of efforts had been devoted
to unify the general theory of relativity (describing the gravitational
interaction) with quantum mechanics (describing the microscopic
interactions of the elementary particles). Yet the conventional
theory of quantum mechanics, based on unitarity and symmetries,
is contradictory with general relativity 
in many respects, $e.g.$ , formally infinite zero point
energy associated with canonical quantization scheme, ultraviolet divergent 
energy density associated with vacuum fluctuations [1] for collapsing
physical systems, unitarity
violation for black hole evolution. 

To unify these two descriptions of nature we can procced
along two directions [2] :

For the Euclide school space-time geometry is an abstract concept
which exist irrespective of matter fields. For the non-Euclide
school space-time does not exist independent of matter fields. 
Space-time is form of existence of matter and can not
be concieved without matter. This feature is even more transperent
from the facts, among others, that the universe is compact and 
there is no well-defined stress-tensor for
gravity which interwinds matter fields and space-time geometry [3].  

In the context of the general theory of relativity the conflict
between the two schools arise in the following way:

As soon as one derives the geodesic deviation equation from
the principle of equivalence one ``can'' forget the source, ascribing
the relative accelaration  of the nearby geodesics to the space-time
manifold. This, in contradiction to the philosophy of general relativity,  
may lead to think that the space-time
manifold is more fundamental leading to the concept of quantum
gravity irrespective of existence
of the corresponding sources that will produce
the fluctuating geometries (no source indicates no space-time geometry
and observationally, quantum fluctuations in matter fields 
are negligible to produce significant alternations
of the space-time geometry). 

Quantum gravity also led to many space-time geometries which are 
physically non-existent. 
One such example is the extreme Reissner-Nordstrom black hole which cannot
be obtained through any realistic gravitational collapse [ ]. It also
has vanishing Hawking temperature. This may be interpreted
through the fact that Hawking radiation through
pair production near the black hole event horizon is not possible
as the metric do not change its signature across the horizon.

We should keep in mind that the Einsteins equations in the 
regime of its validity determine the space-time geometry:
the geometry of space-time as a whole is determined by corresponding
matter fields described 
either in terms of some classical models
or by a proper quantum theory. This gives
a particular cosmology (the closed or mathematically
more properly compact universe picture) and if we
view cosmology as a whole there is really no test body.

We will now consider some aspects of the black hole space-time
geometry. In the process of gravitational collapse an event horizon,
the black hole event horizon, is formed breaking the global
$CP$ invariance and giving rise to the Kerr-Newman families of
black holes(the no hair theorems).  The black hole event horizon may be
defined as the causal boundary of the set of complete time-like
geodesics which originates at the past time-like infinity and 
terminate at the future time-like infinity
as classically nothing can come out
off the horizon. The black hole space-time
is usually described either in terms of the Schwarzschild coordinate
system or in terms of the Kruskal-Szeckers coordinate system.
In the Schwarzschild coordinate system the black hole event horizon
is a two dimensional fixed point set of the time-like Killing vector
field across which some of the metric components change sign. In the 
Kruskal-Szeckers coordinate system the event horizon is a two
dimensional null surface across which the square of some of the
coordinates change sign. We will now consider the 
non-equivalence of the two coordinate systems in detail.

\section{{Non-equivalence of the Schwarzschild and the
Kruskal-Szekers Coordinate system}}

The Schwarzschild space-time is a Lorentz signature,
static spherically symmetric solution of the Einstein
equations when the Ricci tensor vanishes. This solution
describes the exterior geometry of a static spherically
symmetric star and has been used to verify the predictions
of general relativity for the Solar system.

A space-time is said to be static if there exits a space-like
hypersurface which is orthogonal to the orbits of the
time-like Killing vector field. A space-time is said to be
spherically symmetric if the space-like hypersurfaces
contains $SO(3)$ as a subgroup of the group of isometries.
The orbit spheres of $SO(3)$ are isometric to the unit
two sphere. These features together with the condition
of the asymptotic Newtonian limit give the well-known
Schwarzschild solution in the spherical polar coordinates[3]:

\be
ds^2 = -(1 - 2M/r) dt^2 + (1 - 2M/r)^{-1} dr^2
+ r^2 [{d\theta}^2 + \sin^2{\theta}{d\theta}^2]
\ee

According to the Birkhoff's theorem [19] all spherically symmetric
solutions with $R_{ab} = 0$ are static and the Schwarzschild
space-time is the unique static spherically symmetric 
solution, upto
diffeomorphisims, of the Einstein equations with $R_{ab} = 0$.

The norm of the time-like Killing vector field
and ${(\nabla r)}^a$ in the orthonormal coordinates vanishes
and some of the metric components are not well-behaved at $r = 2M$ 
in the Schwarzschild coordinates. The proper acceleraration
of the constant $r$ observers can be obtained from the
geodesic equations in the Schwarzschild coordinates.
This acceleration, $a = {(1 - 2M/r)^{- 1/2}}{M/r^2}$,
is divergent at the horizon $(r = 2M)$. 

The ill-behavedness of the Schwarzschild coordinates
is not a coordinate singularity like
that of the
spherical polar coordinate system where the azimuthal   
angular coordinate $\phi$ become ambiguous at the poles.
All the ill-behavedness of the Schwarzschild
coordinates at the horizon
originate from that of the space-time metric.
The curvature scalars calculated from
the metric are well-behaved
at the horizon
unlike $r = 0$ where the curvature
scalars diverge. For ordinary stars this metric
singularity  at $r = 2M$ is irrelevant as it is inside the star
and the Schwarzschild solution is not valid in the matter
filled interiors. However it is well-known that 
sufficiently massive stars can undergo gravitational
collapse to form black holes and the metric singularity
at the horizon is important. 
Several coordinate systems had been
introduced to remove the metric singularity and to
extend the Schwarzschild space-time where the Schwarzschild
coordinate system is referred to covering a proper submanifold
of the extended space-time. The metric in these extended
coordinate systems are well-defined every where apart
from the space-time singularity. The most well-known
extension is the Kruskal-Szekers coordinanate system.
In this article we perform a comparative study of these two
coordinate systems and show that they are not diffeomorphically
equivalent.

In this section we will follow the abstract index convension
of Wald [3] and extend its significance in Appendix:A.

According to the theory of relativity if $\phi : M \rightarrow M$
is diffeomorphism then $(M, g_{ab})$ and $(M, \phi^* g_{ab})$
represent the same physical space-time. Let a coordinate
system ${x^\mu}$ cover a neighborhood $U$ of a point $p$ and
a coordinate
system ${y^\nu}$ cover a neighborhood $V$ of the point $\phi(p)$.
Now we may use $\phi$ to define a new coordinate system ${{x'}^\mu}$ 
in a neighborhood $O = {\phi^{-1}}[V]$ by setting 
${{x'}^\mu} = y^\mu [\phi (q)]$ for $q$ belonging to $O$. We 
may then take the point of view as $\phi$ leaving $p$ and all
tensors at $p$ unchanged but inducing the coordinate transformation
${x^\mu} \rightarrow {{x'}^\mu}$.
For $\phi$ to be a diffeomorphism ${{\partial {{x'}^\mu}} 
\over{\partial x^\nu}}$ should be non-singular [3,15]. According
to this point of view two coordinate system covering a space-time
can be taken to be equivalent if the corresponding transformation
coefficients are not singular in their common domain of definition
otherwise an arbitrary smooth function defined in one coordinate
system may not remain smooth in the other coordinate system.

To extend the Schwarzschild coordinate system one considers
the two dimensional $r-t$ part:

\be
ds^2 = -(1 - 2M/r) dt^2 + (1 - 2M/r)^{-1} dr^2
\ee

The Regge-Wheeler coordinate system is defined through
the null-geodesics and is given by:

\be
r_* = r + 2M ln (r/2M - 1)
\ee

in this coordinate $r \rightarrow 2M$ corresponds to
$r_* \rightarrow -\infty$. The null coordinates are defined
as:

\be
u = t - r_* , ~~v = t + r_*
\ee

A regular metric is obtained through the following
transformation,

\be
U = -e^{-u/4M} , ~~V = e^{v/4M}
\ee

The metric in these coordinates becomes:

\be
ds^2 = -{{32M^3 e^{-r/2M}} \over {r}}dUdV
\ee

As there is no longer a coordinate singularity at $r = 2M$ 
(i.e at $U = 0$ or $V = 0$) one extends the Schwarzschild
solution by allowing $U,V$ to take all possible values.
However the transformation coefficients 
$dU/dr = -d[{(r/2M - 1)^{1/2} e^{-{(t - r) \over{4M}}}}]/dr$ and
$dV/dr = d[{(r/2M - 1)^{1/2} e^{{(t + r) \over{4M}}}}]/dr$
are singular at $r = 2M$ and the extension is not
diffeomorphically equivalent. Consequently as discussed
at the beginning of this section the Schwarzschild
coordinate system and the $(U,V)$ coordinate system
do not represent physically the same space-time manifold.
Consequently,
according to Birkoff's theorem,
the space-time represented by the $(U,V,\theta,\phi)$
coordinate system
is not a solution of the Einstein equations for a spherically
symmetric black hole.

Similar discussions are valid for the Kruskal-Szekers
coordinate transformations which are obtained through the
following transformations:

\be
T = (U + V)/2 , ~~X = (V - U)/2
\ee

and the metric becomes,

\be
ds^2 = {{32M^3 e^{-r/2M}} \over {r}}(-dT^2 + dX^2)
+ r^2{({d\theta}^2 + \sin^2{\theta}{d\phi}^2)}
\ee

The relation between the $(T,X)$ and the $(t,r)$ coordinates
are well known and in the physical regions of interests are
given by [4],

\be
X = (r/2M - 1)^{1/2} e^{r/4M} {\cosh(t/4M)}
\ee

\be
T = (r/2M - 1)^{1/2} e^{r/4M} {\sinh(t/4M)}
\ee

valid for $r ~> ~2M$, and

\be
T = (1 - r/2M)^{1/2} e^{r/4M} {\cosh(t/4M)}
\ee

\be
X = (1 - r/2M)^{1/2} e^{r/4M} {\sinh(t/4M)}
\ee

valid for $r ~< ~2M$.

Again the transformation coefficients are not defined
on the horizon and the Kruskal-Szekers coordinates
do not give a proper 
diffeomorphic extension of the Schwarzschild coordinate
system. Hence the Kruskal-Szekeres coordinates is not
a solution of the Einsteins equations for a spherically 
symmetric black hole

The Kruskal-Szekers coordinate system had been introduced
to eliminate a particular singular function (the metric
components) in the Schwarzschild coordinate system through
a singular coordinate transformation. This does not ensure
that all singular tensors can be made regular in the new
coordinate system and also tensors which are regular in the
$(t,r)$ coordinates can become singular in the $(T,R)$
coordinates. To illustrate these features we consider
the implicit relations between the two coordinate systems [1]:

\be
(r/2M - 1)e^{r/2M} = X^2 - T^2
\ee

\be
{t \over 2M} = ln({{T + X} \over {X - T}})
\ee
 
The horizon in this coordinates are defined as $X = \pm T$. 

Firstly the proper acceleration of the curves in Kruskale-Szecker's
coordinate system which correspond to the constant $r$ observers
in the Schwarzschild coordinate system
is given by $a = (X^2 - T^2)^{-{1/2}}[e^{r/2M}{M/r^2}]$. This is
also divergent on the horizon.

Secondly we consider the vector 
$({{dR} \over {ds}})^a$, $R^{'a}$,
the proper rate of
change of the curvature scalar $R$ obtained from ${(dR)^a}$ and 
the proper distance $ds$ 
[i.e, the vector $({{dR} \over {ds}})({{\partial} \over {\partial{\bf r}}})$,  
 Appendix:A].
The norm of this vector in the Schwarzschild coordinate
system is $(dR/dr)^2$ and is finite on the horizon. Whereas the
corresponding quantity in the $(T,X)$ coordinates can be
obtained from the following relations
[apart from normalizing factors: $({{\partial X} \over {\partial s}}),
({{\partial T} \over {\partial s}}) = 
{[{{re^{r/2M}} \over {32M^3}}]^{1/2}}$ :

\be
{dR \over dX} = {{\partial R}\over {\partial r}}
{{\partial r}\over {\partial X}} , ~~{dR \over dT} = 
{{\partial R}\over {\partial r}}
{{\partial r}\over {\partial T}} 
\ee

and from equ.(13),

\be
{{\partial r}\over {\partial X}} = {{8 M^2 X e^{-{r/2M}}}
\over{r}}, 
~~{{\partial r}\over {\partial T}} = -{{8 M^2 T e^{-{r/2M}}}
\over{r}} 
\ee

and we have $|{R^{'a}}_{KS}|^2 = 
{{64 M^4 e^{- r/M}} \over {r^2}}{({{\partial R}\over {\partial r}})^2}
[X^2 - T^2] = 0$ on the horizon although
the $r$ -dependent multiplying factor in front of the
Kruskal-Szecker's metric is finite at $r = 2M$. 

The unit space-like normal vector
to the $r = constant$ surfaces, which can be defined apart
from $r = 0$, $k^a = {({{dr}\over{ds}})^a}$ has unit norm ($k^a k_a = 1$)
on $r = 2M$ although $k^a \rightarrow 0$ 
as $r \rightarrow 2M$  which for an outside
observer ($r ~> ~2M$) may be interpreted as nothing can propagate
radially outward at $r = 2M$, consistent with
the divergent acceleration for a radially infalling particle.
Also no combination of the unit time-like normal and the
unit space-like normal to the $r ~= ~const.$ surfaces are possible
whose norm is zero on the horizon but finite for $r ~> ~2M$.

For two metric spaces the definitions of continuity
is as follows [16]:

Let $(S, d_S)$ and $(T, d_T)$ be metric spaces and let
$f: S \rightarrow T$ be a function from $S$ to $T$.
The function $f$ is said to be continuous at a point
$p$ in $S$ if for every infinitesemal $\epsilon > 0$
there is an infintsemal $\delta > 0$ such that 

\be
{d_T}{[f(x),f(p)]} < \epsilon, ~~whenever ~{d_S}{[x,p]} < \delta.
\ee

If $f$ is continuous at every point of $S$ then $f$ is
continuous on $S$.
 
The definition is in accordance with the intuitive idea
that points close to $p$ are mapped by $f$ into points
closed to $f(p)$.
From equn.(13),(14) we have,

\be
|dt|_{Sch} = {{X} \over {(X^2 - T^2)^{1/2}}}{|dT|_{KS}},
~~|dt|_{Sch} = -{{T} \over {(X^2 - T^2)^{1/2}}}{|dX|_{KS}}
\ee

and,

\be
|dr|_{Sch} = {{X} \over {(X^2 - T^2)^{1/2}}}{|dX|_{KS}},
~~|dr|_{Sch} = -{{T} \over {(X^2 - T^2)^{1/2}}}{|dT|_{KS}}
\ee
 
where $|{~~}|$ denotes the norm in the respective coordinate
systems and we find that the coordinate transformation, 
$(t,r) \rightarrow (T,X)$ is not continuous on the horizon
as the multiplicative factors diverge on the horizon $(X = \pm T)$.
Consequently the coordinate transformation $(t,r) \rightarrow (T,X)$
is not a homeomorphism and the two coordinate systems
do not topologically represent the same space-time manifolds [3,17].
Hence we show that that the
Kruskal-Szekers coordinate system is not a proper 
extension of the Schwarzschild cooedinate system
and it is not a solution of the Einsteins equation
for spherically symmetric black hole. We conclude
this discussion with the following note:

For any coordinate system we have,

\be
{g'}_{\mu \nu} = {{\partial {x^ \rho}}\over {\partial {{x'}^ \mu}}} 
{{\partial {x^ \sigma}}\over {\partial {{x'}^ \nu}}} 
{({g_{Sch.}})_{\rho \sigma}}
\ee

Consequently it is not possible to find a
coordinate system with a regular
${g'}_{\mu \nu}$ without absorbing the singularities of
${({g_{Sch.}})_{\rho \sigma}}$ at $r = 2M$ into the
transformation coefficients
${{\partial {x^ \rho}}\over {\partial {{x'}^ \mu}}}$
at $r = 2M$ i.e, without breaking the diffeomorphic
equivalence of the two coordinate systems. Thus,
as also discussed in the preceding sections,
the Kruskal-Szekeres coordinate system 
with a regular metric at the horizon can not be
diffeomorphically equivalent to the Schwarzschild
coordinate system and thus do not represent a static
asymptotically flat solution of the Einsteins
equations representing a blck hole formed out of the 
gravitational collapse of an
uncharged spherically symmetric asymptotically
flat star.
[see also
Appendix:E].

In passing we note that the gravitational collapse to form 
black hole is associated with entropy decrease. The entropy of
a star is propertional to its volume for $r > 2M$ whereas the entropy
becomes propertinal to the area of the horizon, $16 \pi M^2$,
as the star crosses the Schwarzschild radius to form a black hole.

It is not obvious how to describe 
the space-time evolution of the complete
gravitational collapse of matter fields as a
whole in terms of time-like
curves as, for a Schwarzschild observer,
the time-like curves suffer a discontinuity across the horizon
and become space-like inside the
black hole event horizon.
It is welknown
that expressed in terms of the Schwarzschild coordinates
the black hole event horizon has profound impact on the quantum
description of matter fields and black hole evaporation through
Hawking radiation makes the space-time dynamic. Also Hamiltonian
evolution of matter fields break down on the fixed point
sets of the time-like Killing vector field [14]. The canonically
conjugate momentums are not well-defined on the horizon
as will be evident from the lagrangians of the matter
fields.

\section{{Discussion}}

In a gravitational collapse once the collapsing body crosses the
horizon it collapses to form the space-time singularity breaking
the description of space-time in terms of continuous manifolds
and the local symmetries.
We can only characterize the presence of of the space-time singularity
in a diffeomorphism invariant way, in terms of the curvature invariants
along the space-time curves which cross the
event horizon and necessarily terminate along the space-time
singularity. The formation of black hole event horizon
can be characterized through the formation of trapped surfaces. 
The gravitational collapse and the cosmological
evolution are the only two processes in nature through which a
three dimensional physical system collapses to zero dimension
(forming the space-time singularity). Here through zero dimension
we mean a point or a collection of points. We will illustrate
this aspect in Appendix:B.

Einsteins equations break down at the space-time singularity.
This is something similar to electrodynamics. We can determine the
electric field for a point charge using
the Maxwells equations. But the field strength diverges and classical
electrodynamics break down at the point charge (the corresponding
quantum theory QED is not a resolution to this problem. It has
its troubles associated with the point-like interaction terms.
However experimental observations confirm that all the elementary
particles are of finite volume).
The formation of space-time singularity is 
associated with finite volume to zero dimension transition
for the corresponding collapsing body and the 
richest structure that we can attribute to zero dimension 
is that of an analogue of (compact) three dimensional generalization of the
Cantor set [5,6] provided we generalize the description
of the collapsing matter field through a proper quantum theory 
[a generalization of the Pauli exclusion principle].

There are two ways that one can reach zero dimension from finite
volume breaking the continuous topology of space-time manifold.
One is through the point contraction mapping which 
requires an infinite number of
iterations which, together with the discussions in
Appendix:B, is in accordance with the fact that time is
a continuous parameter.
The other one is through the formation of 
an analogue of the Cantor
set (or any other discrete manifold with different
cardinality) 
in which case the underlying physical processe 
to achieve zero dimension may be discontinuous. 
A discrete manifold may not always form a normed vector space, e.g,
the set of points $(n ~+ ~x)$ on 
the real line, where $n$
is an integer and $x$ is a fractional number, can not
form a normed vector space as the difference between two
points do not belong to the set. Also it is not physically obvious
to talk of causal structure, defined through propagation of signals,
in a discrete manifold unless the manifold is space-like
and frozen in time (which is defined through physical processes).
As discussed earlier according to the General Theory of Relativity
charges associated with space-time transformation symmetries
are global properties of a continuous space-time manifold
as a whole whereas a 
manifold without continuous topology can only have space-time
independent charge.

To describe cosmological evolution and black hole evolution we will
have to generalize and geometrize conventional quantum mechanics
in a suitable way. In these respects
the principal aspects to be critically studied, as will be discussed
later in this article, towards unifying the general
theory of relativity and conventional quantum mechanics is the concept
of diffeomorphism invariance associated with the general theory
of relativity and unitarity associated with the conventional
quantum mechanics.

The facts that the continuous topology of space-time break down
at the space-time singularity [Appendix:B]
(indicating no well-defined observables associated with spatial
transformations for the cosmologically evolving or collapsing matter fields
in the near zero dimension region) and 
that nature choses a particular
cosmology lead us
to conclude that diffeomorphism invariance (which for large scale
structure of space-time is equivalent to invariance under
coordinate transformations) is not of so important
(as it is for solar system microscopic physics)
concern for the corresponding physical laws.
Rather the fact Schwarzschild coordinates and the Kruscal-
Szekers coordinates are not diffeomorphically equivalent indicates 
that an appropriate
choice of a suitable coordinate system is most important.
However we can express the generalized
quantum theory in covariant form. This will help us to compare
the generalized quantum theory with solar system quantum physics where
the physical laws are invariant under inertial transformations
and are formulated in a covariant way under the corresponding
coordinate transformations.

To generalize the conventional quantum mechanics we should take
into account the following important aspects:

(1) Special relativity made the concept of size for ordinary objects
a relative one.
The strong curvature effects near the space-time singularity will
spoil the concept of dimension for the elementary particles
forming the space-time singularity.

(2) Quantum mechanics is the mechanics of quantum states
which do not exist independent of their realizations at least
in principle, i.e, through interactions with other quantum states.
For solar system microscopic physics the fact that the elementary
particles or bound states formed by them are of finite
volume has to be considered in the corresponding 
quantum state description as long as the 
continuous structure of the space-time manifold
holds. 

(3) Every measurement process through state reduction is a non-unitary
operation on the space of quantum states [7]. Near the space-time singularity
the strong carvature will destroy the description of matter fields
in terms of a unitary quantum theory.

As far as the 
cosmological evolution is concerned (zero dimension to finite volume
and finite volume to zero dimension transitions) no observer physics
is the exact description of the evolution of the 
universe in the near zero volume
region. A proper generalization of quantum mechanics may be non-unitary in the 
sence that the evolutions of the possible quantum states 
(if the space-time description of matter is given by 
a particular family of space-time curves representing
possible particle trajectories [5]) representing
the collapsing physical systems may be non-unitary.

However, it is obvious that the collapsing physical system
(which is a bound system through gravitational interaction)
collapsses to zero dimension violating the conventional
quantum mechanics based on the uncertainity principle
(e.g, the electrons in an atom obeying the quantum
mechanical principles do not collapse on the positively charged
nucleas)
and follow the
deterministic laws of general relativity.
 
In the context of the above
discussions an important contradiction between 
the general theory of relativity (describing
the gravitational interaction) with conventional quantum mechanics 
(describing the microscopic
interactions of the elementary particles)
is the following:

Positivity of the energy momentum stress-tensor together with the
general theory of relativity leads to 
gravitational collapses [8]
and space-time singularities [9]
where a three
dimensional physical system collapses to zero dimension (breaking the
continuous space-time topology)
whereas positivity of the energy-momentum tensor together with the
canonical commutation relations lead 
to the Pauli exclusion principle (unless one introduces
additional structures about the space-time singularity). 
 
We, living beings, are characterized by the fact that we can control
some terrestial processes. But we can neither change the physical laws
nor the cosmological evolution. Many descriptions we had made are
either incomplete (unitary quantum mechanics) or
approximations (point particles for microscopic physics). 
The discussions in the preceding paragraph
together with the facts that black holes contain
no scalar hair [11],
that there is no physical explanation of only recombination for
the virtual particles (which ``interpret'' real effects
as in the Casimir effect) 
to form loops in quantum field theory 
[Physically, in vacuum, even in Feynmann's summing over path scheme
it is not obvious why particle pairs produced at one space-time
point will only recombine at another space-time point. The
otherwise should give abandunces of particles and
antiparticles. We will discuss this issue in some details
in Appendix:C.]
and 
that the universality of the minimum uncertainity relations are
lost in the gravitational collapses and 
are questionable in the solar system
microscopic physics [12] lead to conclude that the proper
avenue towards unifying these two theories and thereby explaining
the cosmological evolution completely will be understanding the
space-time singularity and extending the conventional quantum theory 
as the position-momentum Canonical commutation relations
are in accordance with the corresponding minimum
uncertainity relations [13].

\section{{Appendix:A}}

We can obtain the one-form $({\bf{d\phi}})_a$ from a zero-form
(a scalar field) $\phi$ in an explicit coordinate variables
notation:

\be
({\bf{d\phi}})_a = {\sum{{{\partial \phi} \over {\partial x^{\mu}}}
[{\bf{(dx)}}_a]^{\mu}}}
\ee

here the range of the summation is the dimension of space-time
and it does not represent an infinitsemal change in $\phi$.
When expressed in a particular coordinate basis ${[{\bf{(dx)}}_a]^{\mu}}$
will be just ${\bf{dx}}^\mu$,
a coordinate one-form, and the $\mu$ -th component of $(d\phi)_a$
is ${{\partial \phi} \over {\partial x^{\mu}}}$
as an arbitrary tensor $T$, in its operator form,
represented in a coordinate
basis can be expressed as:

\be
{\bf T} = {{T^{\alpha \beta ...}}_{\mu \nu ...}}
{({{{\bf{\partial}}} \over {{\bf{{\partial} x}}^\alpha}})}
{({{{\bf{\partial}}} \over {{\bf{{\partial} x}}^\beta}})}...
({\bf{dx}}^\mu)({\bf{dx}}^\nu)...
\ee 

Here
${({{{\bf{\partial}}} \over {{\bf{{\partial} x}}^\alpha}})},
{({{{\bf{\partial}}} \over {{\bf{{\partial} x}}^\beta}})}$
are coordinate unit vecrors and 
${\bf{dx}}^{\mu,\nu}$ are coordinate one-forms.

In terms of a coordinate basis the covariant
d Alembaratian operator can be obtained from the invariant:

\be
{({\bf{d\phi}})^a}{({\bf{d\phi}})_a} = {\sum \sum{{g}^{\mu \nu}}
{{{\partial \phi} \over {\partial x^{\mu}}}
{{\partial \phi} \over {\partial x^{\nu}}}}}
\ee

here the explicit summations are again over $\mu, \nu$ with the ranges same
as above. 
When expressed explicitly in a coordinate basis
the Lagrangian density of a massless scalar field is given
by: ${g_{\mu \nu}}{{\partial^{\mu}{\phi}}
{\partial^{\nu}{\phi}}}     
                       = {g^{\mu \nu}}{{\partial_{\mu}{\phi}}
{\partial_{\nu}{\phi}}}$.
The infinitesemal change $\delta \phi$ of the scalar field
$\phi$ can be interpreted as the scalar product of the one
form $({\bf{d\phi}})_{a}$ and the infinitsemal vector line elements.

\section{{Appendix:B}}

In this section we will briefly discuss the relationship
between the fundamental properties of the elementary
particles and space-time geometry. We will show that
a line-element is not just a collection of points.
Line-elements, area-elements and volume-elements
are as fundamental as points. We also demonstrate
that three spatial dimentions is in accordace with 
the finite volume and spin of the elementary
particles to give unique dynamics. We will also discuss
the consequences of these aspects for geometry
and coordinatization. 

A point is a dimensionless object. A line is a one dimensional
object. Two lines intersect at a point. If two lines intersect
each other at every point they are said to be coincident.
Two non-coincident lines are said to be parallel if they never
intersect each other. A plane is a two dimensional object and
any point on the plane can be 
characterized through the choice of two lines on the plane and
constructing a coordinate system in the usual way. There are
two possible motions on a plane, translations and rotations.
Rotations can be uniquly characterized only through their
magnitude and the unique
normal to the plane. In three dimension any infinitesemal
rotation can be taken to take place on a plane uniquly
characterized through its unique normal which is contained
within the three dimensional spatial geometry. Higher
dimensions greater than three will spoil the uniquness
of the normal to the plane of rotation. 
Thus three dimensional spatial geometry is self
-complete both geometrically and 
to describe the dynamics of matter particles uniquly.
Also the concept of orientation with a proper
convention is an essential aspect
to formulate the laws of Classical Electrodynamics 
consistent with the energy conservation law. The
fact that in general, apart from the case
when the plane of the loop is perpendicular
to the magnetic field, 
a time-varying
magnetic field always induces an electric current
in a closed loop 
obeying Faraday's
law indicate that spacial geometry should be of
three dimension.

A line is not just a collection of points as a collection
of dimensionless objects (points), however large may be in
cardinality, cannot give a dimensionful object. For example
the total length of the deleted intervals to form
the Cantor set (a collection of discrete points) 
is 1 although the cardinality of the Cantor
set is same as the original 
unit length interval [5] considered, conventionally, as 
a collection of points. On a line ``two points separated 
by an interval (finite or infinitesemal)'' or ``two points
coincident'' have meaning but ``two points adjuscent''
is not defined. We have to introduce line
intervals, finite or infinitesemal, as fundamental mathematical
entities to form lines giving them the continuous topology. 
We illustrate this feature in the context
of coordinatization of the real line.     

An arbtrary real rational number may expressed as
$r = n.{x_1}{x_2}{x_3}............{x_p}$, where
$n$ is an integer, 
$x_i = 0,...,9$ for $i < p$,
$x_j = 0$ for $j > p$,
$p$ may be arbitrarily large but finite and
${x_p} \neq 0$, i.e., the sequence of the
decimal places is finite.
Whereas an irrational number is given by:
$ir = n.{x_1}{x_2}{x_3}............$
and the sequence of the decimal places runs to infinity.
Here the definitions differ from the conventionals.
We will consider $r,ir ~> ~0$ in the following section.

It is easy to find that between any two rational numbers
there are infinite number of irrational numbers as any 
irrational number
(with $x_i$ same as that of $r$ for all $i < p$) 
whose $p$ -th decimal number is equal
to $x_p$ 
is greater than $r$ and any irrational number
(again with $x_i$ same as that of $r$ for all $i < p$)
whose $p$ -th decimal number is equal to
$(x_p - 1)$ is lower than $r$. 
Similar arguments as above and discussed below, are valid for 
two irrational numbers as
the sequence of decimal places for an irrational
number is infinite. 

We now prove the following proposition:

On a line the concept "two points adjuscent" is not well-defined. 
However close two points may be we will always find
a point lying between them. Thus on a line only the
statements 
"two points separated by an interval" or "two points
coincident" are well-defined and line-intervals are
fundamental gometric objects.

If someone say that the two points on the real line
represented by a rational number $r (\neq ~an ~integer)$ and 
an irrational
number $ir$ are adjuscent; we can always find an irrational
number, following the construction
in the paragraph before the last one with any $x_i ~{(i > p)}$
replacing $x_p$, which is
greater than or lower than $ir$ and is nearer to $r$.

If $r$ is an integer $(> 0)$ and 
$ir = r.{x_1}{x_2}{x_3}............$, with all but
$x_p ~(= 1)$ are zero for $p \rightarrow \infty$, the whole
region infinity on $R^1$ can be used through a
mapping of the decimal position index $p$ of $ir$ for 
$p \rightarrow \infty$
(with only ${x_p}$ nonzero, $= 1$) to the integers in the region
infinity of $R^1$ to construct an irrational number less than
$ir$ and closer to $r$
[i.e, the number of decimal places having value zero
before ${x_p} = 1$ for $p \rightarrow \infty$ can be
increased indefinitly]. 
Similar will be the case
for $ir = (r - 1).999......$
,with all ${x_i} = 9$ but ${x_j} = 8$
for $j \rightarrow \infty$ and $j > i$, 
to construct an irrational number
greater than
$ir$ and closer to $r$
through a
mapping of the decimal position index $j$ of $ir$ for 
$j \rightarrow \infty$
(with all ${x_i} = 9$, ${x_j} = 8$ for $j > i$) 
to the integers in the region
infinity of $R^1$.
[i.e, the number of decimal places
having value 9 before ${x_j} = 8$ for $j \rightarrow \infty$
can be increased indefinitly].

Similar as above will be the arguments with $(n \geq 0)$
$r = n.{x_1}{x_2}{x_3}............{x_p}$,
$ir = n.{x_1}{x_2}{x_3}............$ 
(${x_i}$ same for both for $1 \leq i \leq p$)
with
all $x_i ~{(i > p)}$ but $x_j ~{(= 1)}$ are zero for
$j \rightarrow \infty$ and for
$ir = n.{x_1}{x_2}{x_3}............{y_p}999....$ with
the $p$ -th decimal place $y_p = x_p - 1$
and all ${x_i} ~(i > p)$ but ${x_j} ~(= 8, j > i)$ for 
$j \rightarrow \infty$ are equal to 9.

For two irrational numbers $(ir)_1 , (ir)_2 $ we can
always recoordinatize $R^1$ so that $(ir)_1 $ become
a rational number and whether $(ir)_2 $ is a rational
number or not the arguments in the above paragraphs
show that there are non-denumerably infinite number of points between
$(ir)_1 $ and $(ir)_2 $.

For $r ~\leq ~0$ the corresponding arguments to prove that ``two
points on $R^1$ are adjuscent'' is not defined are
very similar as in the preceding paragraphs.
 
The above arguments lead us to consider line interval as the
fundamental geometrical entity to form a line  
and the real line is not just
an array of ordered points. Irrational numbers,
as defined in this section, demonstrate
the continuous topology (existence of neighbourhoods)
of the real line.

Similarly, as we will illustrate later in
this section, we cannot obtain
a two dimensional plane from a collection of lines
and a three dimensional hyperplane from a collection
of planes.
Thus one, two and three dimensional intervals are fundamental
geometrical entities to have one, two and three dimensional
geometries and give the continuous topology. One can not
obtain points without spoiling the continuous topology.
The three dimensional spatial geometry of the Universe is realized
through the finite three dimensional volume of the
fundamental particles and the finite three dimensional
volumes of the fundamental particles can only lead to
three space dimensions for the Universe.

In brief following the principles of
the General Theory of Relativity: a consistent and unique dynamics of the
the Universe 
is realized through the finite three dimensional
volumes and spin, an intrinsic property distinguishing
spatial directions and orientations, 
of the elementary particles and the polarizaton
and extended nature of fields. Also as
discussed earlier the concept of 
orientation is an esential aspect to formulate the
laws of Classical Electrodynamics [24] which require
that spacial geometry should be of three dimansion.  
 
We note as the real line
cannot be defined as a collection of points we have
to introduce the concept of collection of one-dimensional
line elements as a fundamental mathematical entity
which, at its most primitive level, can be characterized
into four classes:

(i)one-dimensional line elements without boundaries

(ii)one-dimensional line elements with one boundaries

(iii)one-dimensional line elements with two boundaries

(iv)one-dimensional line elements with the two boundaries
identified (e.g, a circle)

We can introduce additional structures like length
for the elements belonging to this collection.
For example we can define the length
of a circle (perimeter) and introduce
the concept of its radius as the perimeter over
$2\pi$. The values of the radius form
an abstract one-dimensional line element and 
,as shown earlier, the notion of two everywhere adjuscent circles
(i.e, two circles with adjuscent values of radii with
everything else remaining the same)
is not defined. This feature again leads 
us to conclude that we cannot
obtain a two-dimensional disk from a collection of
one-dimensional circles. We have to introduce the
two dimensional circular strips as fundamental mathmatical
objects to construct a two-dimensional disk in
the above way. 
Similar arguments for two dimensional spheres
(where the radius is now defined as the positive
square root of the area of the two-dimensional sphere 
over $4\pi$) lead us to conclude that we can not have
a three dimensional volume element from a collection
of two dimensional spheres and we have to consider
three dimensional volume elements as fundamental
mathmetical entities. These discussions can be extended
to higher dimensions.

We now discuss a few aspects regarding coordinatization.
As far as the number system is concerned
the above discussions prove that for any given number
it is not possible to define the concepts of the
immediate previous number or the next number; only
the concepts of earlier or later numbers are well-defined.
Consequenly the number system with it's conventional
interpretation as one-to-one correspondence between
numbers and points cannot cover a one-dimensional line.
We can illustrate this again through the Cantor set.
After we have removed all the intervals, whose total
length is same as the original interval, the rmaining
set of points forming the Cantor set have 
continuuam cardinality,i.e, the elements of the Cantor
set, through the very definition of cardinality, 
can be put into a one-to-one correspondence with
the Real number system again.
To summerize, the number system with it's conventional
interpretation as one-to-one correspondence between
numbers and points can only characterize intervals. 
However the facts that elementary particles are of
finite volume, the non-localized character of fields,
the kinematic equivalence of space and time through
the principles of Relativity (valid apart from the space-time
singularity) and the dynamical character of the Universe
indicate that space-time intervals 
are fundamental aspects and thereby the
conventional coordinatization
(there is no prblem, if required, in coordinatizing a particular
point as origin, even it may be so that 
at the origin the spce-time is not well-defined) 
scheme defined through the
concepts of neighbourhoods works well 
as far as physical evolutions are
concerned.

We conclude this section with a few discussions:

Firstly let us consider two sets each
containing a single object: a closed line-element. The
line-elements are intersecting but non-coincident
everywhere. The intersection of these two sets
is a set containing points which
are fundamentally different
from line elements: we can not say that a point
is a line-element with one point as the
arguments in this section prove that a line-element
is not a collection of points. 
Also as a line-element is not just a collection
of points a set-theoretic union
to form a closed line-element out of
a set of half-open line-elements, a set of
open line-elements and a set of points is not well-defined
as all the above mentioned sets are geometrically and even
topologically different. Similar discussions remain
valid in two dimensions when two area elements intersect
along a line-element and in three dimensions when two
closed volume-element intersect along a common boundary.
These problems can be solved if we construct a universal
set containing all possible geometric-elements including
points.

Secondly, the discussions in this section will also have
some profound
significances in Mathematical Analysis. 

To illustrate we make some comments regarding 
the first and second countability of $R^1$ discussed in Appendix:A
of Wald [3].
A topological space $X$ is first-countable if for every
point $p$ belonging to $X$ there is a countable collection of
open sets such that every open neighbourhood of $p$
contains at least one member of the collection. Whereas
$X$ is second countable if there is a countable collection of
open sets such that every open set can be expressed
as an union of open sets from this family. For $R^1$,
the open balls with rational radii centered on points
with rational coordinates form such a countable collection open sets.

When defined in the conventional way there are infinite number
of irrational numbers between two rational numbers [6].
As far as first-countability is concerned, 
an open set $V$ centered on a rational number with deleted peremeter
on one of these neighbouring irrational number 
cannot contain
open balls with rational radii centered on points with
rational coordinates. 

As far as second-countability is concerned the above-mentioned
open set
can not have locally finite
subcover as the radii of the set of open intervals
centered on the given rational number with deleted perimeter
less than that of $V$
form a line-interval and as is proved in this
article the corresponding cardinality is not
well-defined
Similar discussions can be extended to $R^n$.

\section{{Appendix:C}}

Quantum field theory is the quantum theory of fields. It
gives the dynamics of fields, the quantum probability
amplitudes of
creation and annihilation of particles, in contrast to quantum
mechanics which gives the dynamics of the particles themselves
obeying quantum principle. For the same
boundary conditions these two descriptions match in the form
of their kinemetic solutions. Only for the free particle
boundary condition the conventional interpretation of propagators
in Q.F.T as giving the probability of particle propagation
is in accordance with reality as the quantum probabilities
are nowhere vanishing in both the theories. 
For microscopic particle physics experiments the free
particle boundary condition is a good approximation in
practice but ideally the field $\phi$ 
(or the quantum mechanical wave function $\psi$) is
spatially confined within the experimental apparatus. 
In one dimension it is meaningless
to say that a particle is propagating from one point to another
if the probability of finding (or creation of) the particle
is vanishing at some intermediate points. 

For loops in one dimension, the momentum-space calculations
give the probability that a pair of particles with given
four-momentums are created at one space-time point and a pair
of particles are annihilated at another space-time point
with the same four-momentums. This feature is transparent
if one consider all possible space-time particle trajectories to form
loops in one dimension which cannot be possible without the possible 
space-time particle-anitiparticle (originated with given four-momentums
and annihilated with the same four-momentums) trajectories
crossing each other at least once in between 
any two given space-time 
points. Similar feature will be
apparent if one interpret the loop as a particle encircling
between any two given space-time points with the four-momentums
at these two given space-time points remaining the same.

Let us illustrate this feature following the standard literature.
We consider situations where free particle appeoximation
hold. 
In vacuum at a given space-time point $x$ (in one dimension), the particle-
antiparticle pair production probabilities with two-momentums $p_1,p_2$ are
$|\exp(-ip_1.x_1)|^2$ and $|\exp(-ip_2.x_2)|^2$ respectivly 
apart from normalizing factors. Once produced the 
quantum mechanically allowed stationary state position-space wave functions
that are available to
the particle-antiparticle pairs are
$\psi_P (x) = \exp(-ip_1.x_1)$ 
and $\psi_{AP} (x) = \exp(-ip_2.x_2)$ respectivly where
$x$ denotes space-time points. The quantum mechanical 
joint probability 
that the particles produced  at
$x_1, x_2= 0$
with two-momentums $p_1$ and $p_2$ can 
again coincide at a space-time point between $x$
and $x + dx$ is:

\be
P_x (p_1,p_2) =  N_1{{[dx]^2}\over {T^2 L^2}} 
\ee

here $N_1$ is the relative pair creation probability
at the space-time point $x = 0$.

$P_x (p_1,p_2)$ is independent of $x, (p_1, p_2)$ and
$\rightarrow 0$ as for free particle approximation 
$L \rightarrow \infty$ although the total probability
of coincidence is unity when integrated over all
space-time points. Hence quantum mechanically, numerous amount
of particle-antiparticle pairs should be observed in
any microscopic experiment performed during finite time-interval
if there would have been
spontaneous pair creations in vacuum.

In passing we note that a space-time formed out
of loops in vacuum, closed time-like curves,
as the source can not
have an intrinsic time orientation in contrary to what is realized
in nature.

We next note that in non-relativistic quantum mechanics
the total joint probability that two distinguishable
particles with energy $E$ 
and momentums
$(k,-k; {k = {2n\pi \over L}},|n| >> 1)$
can coinside at some point $x$ (here $x$ is position only) is:

\be
P(all) = {4\over L^2}\int \int{[\int{\sin^2(kx_1) \delta(x_1 - x_2)
\sin^2(kx_2)}dx_2]}[dx_1]^2
\ee

here the integrals are performed over the interval $[-L/2 , L/2]$.
This expression turns out to be unphysical as the
corresponding probability
turns out to be unphysical $[P(all) = 3/2]$. Whereas classical 
mechanically the maximum value of  
$P(all)$ can be nearly unity when 
the particles 
suffer impulsive elastic collisions  
to stick together and come to rest at some point $x_0$ (the
corresponding quantum mechanical probability density should 
have been $\delta^2(x - x_0)$).

Similar features as discussed in this
section in the contexts of
equ.(24) and equ.(25) will appear in three dimensions.

In semiconductor physics, as charge carriers holes are 
fictituous objects introduced for simplifications. In
reality, quantum mechanically in a p-type semiconductor
the motion of holes are out of the movements of the 
valence band or the acceptor level electrons. What
is the proper explanation of the polarity of
the Hall potential in a p-type semiconductor?

\section{{Appendix:D}}

In this section we first consider the action for the gravitational
field [20]:

\be
S_g = -{{c^3}\over{16 \pi k}}{\int{G{\sqrt{-g}}}d^4 x}
\ee

where $G_{ab}$ is Einstein's tensor. General Theory of
Relativity interwinds inertial mass (in general
energy-momentum) of matter with space-time
through the principle of equivalence
and the dimension of the coupling constant $k$ 
($k = {6.67 \times 10^{-8}}{cm^3}-{gm^{-1}}-{sec^{-2}}$)  is
completely determined
in terms of only mass, space and time unlike, for example,
in electrodynamics where one have another
fundamental quantity (elctric charge) to determine the dimension
of the coupling constant. This feature is
transparent if we compare the Newtonian-limit of the 
general theory of relativity with the Coulomb's law.
The consequence of this feature is the following:

If we set $c,{h\over{2\pi}} = 1$ the dimension of
$k$ is length-squared ($[l^2]$) and it is no longer possible
to set $k = 1$ as this will make the concept of
space-time dimensions meaningless. Alternatively
we could have set $c,k = 1$ (see  
footnote: page no. 269, [20]) then Planck's
constant become dimensionful ($[l^2]$).

However we can set Boltzmann constant $(k') = 1$ by giving
temperature the dimension of energy. In the reduced
units $c,{h\over{2\pi}},k' = 1$ the gravitational
action becomes:

\be
S_g = -K{\int{G{\sqrt{-g}}}d^4 x}
\ee 

where $K$ have dimension $[l^{-2}]$.

We now conclude our discussion on Quantum
Field Theory. In the rest of this section we will use the convention
that $<\alpha|\alpha>$ gives us probability density.

We first consider non-relativistic quantum mechanics
in one dimension. In position-space the normalized quantum mechanical
wave function $\psi$ gives us the probability amplitude.
$({\psi}^*{\psi}) dx$ gives us the probability of finding
the particle within the infinitesemal length interval
$dx$. For a free particle one adopts the delta function
normalization scheme for the quantum mechanical wave function:

\be
{{{\int_{-\infty}^{\infty}}}{\psi^{*}_{k_1}(x)}{\psi_{k_2}(x)}dx} =
\delta(k_1 - k_2)
\ee

In this equation the left-hand side is dimensionless while
the one-dimensional delta function has dimension of length $[l]$
as is obvious from it's definition:
\be
{{{\int_{-\infty}^{\infty}}}{f(k)\delta(k - k_{0})}dk} = f(k_0)
\ee

for a regular function $f(k)$.

It would be appropriate to replace the free-particle boundary
condition by periodic boundary condition which is a reasonable
approximation in situations where free-particle boundary
conditions hold as for a large length interval the spacing
between the adjuscent values of the momentum allowed by
the periodic boundary condition is negligible.

In the reduced units $(c, {h \over {2\pi}} = 1)$
the action is dimensionless. The action for a complex scalar field
is given by:

\be
I = {\int{L {d^4 x}}}
\ee

where the covariant lagrangian density for a massive field is given 
by,

\be  
L = {{1\over 2}{{\partial \phi^*}\over {\partial x_{\mu}}}
{{\partial \phi}\over {\partial x^{\mu}}}} - {{m^2 \over 2}{\phi^*}^2
{\phi}^2}
\ee

Consequently the dimension of $\phi$ ($\phi^*$) should be inverse of
length $([l]^{-1})$. In the second quantization scheme 
$<\beta|{\phi}|\alpha>$ replaces the classical field [21] and
the expression $<\alpha|{\phi^*}{\phi}|\alpha>$
gives the probability density of creation or annihilation of
particles.  
For free-particle boundary conditions 
the Euclidean-space generating functional
for a real scalar field is given by [22]:

\be
W_{E}[J] = {N_E}{e^{-{S_{E}{[{\phi}_{0},J]}
-{1 \over 2}{Tr {~} ln[({-{\bar{\partial}}_{\mu}
{\bar{\partial}}^{\mu} + m^2 + V''({\phi}_{0})})_1
{\delta(\bar{x_1} - \bar{x_2})}]}}}}
\ee

The terms in the logarithm giving quantum 
corrections are not dimensionless and
the third term is not of the same dimension as of the first
two terms.

For a real scalar field
confined within a finite volume
box with periodic boundary
condition and consistent with the
second quantization scheme we have (equ.3.28,[10]): 

\be
\phi(x) = {1 \over L^{3/2}}{{{\sum}_{(n_1 , n_2 , n_3)}}
{~}{{{\phi^{+}}_{n_1 , n_2 , n_3}}\over {\sqrt{2k^0}}}
{\exp[{{2 \pi i}\over L}(n_{\mu} x^{\mu})]}} +
{1 \over L^{3/2}}{{{\sum}_{(n_1 , n_2 , n_3)}}
{~}{{{\phi^{-}}_{n_1 , n_2 , n_3}}\over {\sqrt{2k^0}}}
{\exp[-{{2 \pi i}\over L}(n_{\mu} x^{\mu})]}} 
\ee

in a relativistic theory
a covariant normalization using four volume would be appropriate
and the normalizing factor should have dimension $[l^{-2}]$. 
Clearly the dimension of $\phi$ $([l^{-{3 \over 2}}])$ 
in equ.(33) is no longer 
$[l^{-1}]$ and equ.(31) no longer gives us a lagrangian density.

In other words the dimension of the 
scalar field $\phi$ as required from
the action determining it's space-time evolution
does not mach with the dimension required in the second
quantization scheme in order that one can interprete
$<\alpha|{\phi^*}{\phi}|\alpha>$ as giving the probability density
of creation or annihilation of particles.
One can absorbe the the normalizing factor into the the
fock state by mutiplying it by a factor with dimension
$[l^{-1}]$ as in the second quantization scheme
$<\beta|{\phi}|\alpha>$ replaces the classical field.
This will be in accordance with the 
probabilistic interpretation of the field $\phi$
as we have,

\be
<\alpha|{\phi^*}{\phi}|\alpha>{~} = {~}<\alpha|{\phi^*}|\beta>{{\sum}_{\beta}}
<\beta|{\phi}|\alpha>.
\ee

where the sum is taken over all possible states.

However this will violate the interpretation
of the Fock states as quantum mechanically the normalization of 
probablity density $<\alpha|\alpha>$ in a relativistic
theory requires 
that the each of the normalizing factors for the Fock states, 
whose number depends on
the number of particles present in the Fock state,
should have dimension
$[l^{-2}]$.

We now consider fermions and electromagnetic fields.

The covariant lagrangian
density for each component of
the free fermion fields (e.g, electrons-positrons) which
is formed out of their causal space-time motion is
given by:  

\be  
L = {{1\over 2}{{\partial \psi^*}\over {\partial x_{\mu}}}
{{\partial \psi}\over {\partial x^{\mu}}}} - {{m^2 \over 2}{\psi^*}^2
{\psi}^2}
\ee

and dimension of $\psi$ is again $([l]^{-1})$. After linearization
of the second order partial differential equation satisfied by $\psi$
we get the Dirac equation:

\be
(i {\gamma}^\mu {\partial_{\mu}} - m)\psi = 0
\ee

Hereafter the following lagrangian density 
is used to study Q.E.D:

\be
L' = {\bar \psi}(i {\gamma}^\mu {\partial_{\mu}} - e{\gamma}^\mu{A_\mu}
- m)\psi - {1\over 4}{F_{\mu \nu} (x)}{F^{\mu \nu} (x)}
\ee
 
with the dimensions of $\psi$ as determined above
the dimension of the first part of $L'$ is no longer
that of a Lagrangian density and the action formed
out of it is not dimensionless in the reduced units.
Also the current density, $e{\bar \psi}{\gamma}^\mu{\psi}$
(although the four divergence vanishes),
do not involve any momentum operator and it is not
obvious whether it is possible to have, in any
approximation, the conventional interpretation of current
density as charge-density times velocity from this expression.

The Lagrangian density of a charged scalar
field which is similar to the quadratic
Lagrangian density for Q.E.D is
given by,

\be
L_{T} = L_{f} + L_{em} + L_{int}
\ee

This complete Lagrangian density
for a charged scalar is gauge
invariant only if we take $L_{int}$ to be,

\be 
L_{int} = -{A_\mu}{j^\mu} + {e^2}{A^2}{{\phi}^*}{\phi} 
\ee
 
the second term do not have a transparent interpretation
unless we consider screening effects from classical
electrodynamics similar to the correponding discussions   
given in Appendix:II of this article.

\section{{Appendix: E}}

In this section we will illustrate the discussions in the 
context of equ.(20) in section IV.

The metric of the two-sphere $S^2 (\theta, \phi)$ is given by

\be
ds^2 = {\sqrt{A/4\pi}}(d{\theta}^2 + {\sin^2{\theta}}d{\phi}^2)
\ee

Here $0 \le \theta \le \pi$ and $0 \le \phi \le 2\pi$.
For the unit two-sphere we have,

\be
ds^2 = d{\theta}^2 + {\sin^2{\theta}}d{\phi}^2
\ee

with the ranges of $\theta, \phi$ same as above.
This coodinate system have the following ill-behavedneses:

(i) The coordinate $\phi$ suffers a discontinuity along
some direction from $2\pi$ to 0.

(ii) $\phi$ is degenerate at the poles $\theta = 0, \pi$.
In spherical polar coordinate system $(r, \theta, \phi)$
the point $(r = c, \theta = 0)$
where $c$ is a finite constant is obtained through
identifying any two arbitrary points on a circle
characterized only through distinct values of $\phi$.
Similar construction is valid for the point $(r = c, \theta = \pi)$. 
This will be obvious if we construct a two-sphere $S^{2}(\theta, \phi)$
from a two-dimensional circular strip by identifying
the inner-boundary and the outer-boundary to two distinct
points [see the discussions below eqn.(50) regarding 
the reduction of eqn.(48) to eqn.(50) for points
on the polar axis]. This construction can be generalized to higher
dimension.

(iii) The metric is singular at the poles 
[see the above discussions and App.B (a point can't be obtained
from a one-dimensional line-element without breaking the corresponding
continuous topology)].

When the $S^2 (\theta, \phi)$ can be embedded in
the three dimensional Euclidean space one can introduce
Cartezian coordinate system $(x,y,z)$ through the
coordinate transformation:

\be
x = {\sin \theta}{\cos \phi}, ~~~ y = {\sin \theta}{\sin \phi}, 
~~~ z = {\cos \theta}.
\ee

here $x^2 + y^2 + z^2 = 1$. Although the metric is regular
in the Cartezian coordinate system the transformation
coeffcients $({{\partial{{x'}^\mu}}\over{\partial{{x}^\nu}}})$ 
are singular at the poles and also at some
isolated points on the $x-y$ plane demonstrating the 
discussions below equ.(20). On the other hand to obtain
the Spherical metric, singular at two isolated points,
from the regular Cartezian metric one has to introduce
coordinate transformation with singular transformation
coeffcients.

We can also introduce two homeomorphic stereographic
projections to coordinatize $S^2 (\theta, \phi)$
embedded in $R^3$. The first one is from the North
pole $\theta = 0$ on the
Equator plane to coordinatize the Southern hemisphere
${\pi \over 2} \le \theta \le \pi$. We have,

\be
X = \cot{\theta \over 2}{\cos \phi}, ~~~~ Y = \cot{\theta \over 2}{\sin \phi}
\ee

and this transformation is a homeomorphism at $\theta = \pi$.
The second stereographic transformation is from the South pole
$\theta = \pi$ on the Equator plane to coordinatize the
Northern hemisphere $0 \le \theta \le {\pi \over 2}$. 
We have,

\be
U = \tan{\theta \over 2}{\cos \phi}, ~~~~ V = \tan{\theta \over 2}{\sin \phi}
\ee

and this transformation is a homeomorphism at $\theta = 0$.

The transformation between the $(X,Y)$ and $(U,V)$
coordinate systems is a diffeomorphism at their common domain 
$\theta = {\pi \over 2}$. The metric is also regular
in these coordinate systems.

However the transformation coefficients between $(X,Y)$ and
$(\theta, \phi)$ coordinates are singular at the South pole
$({{\partial{X}}\over{\partial{\phi}}},
{{\partial{Y}}\over{\partial{\phi}}} = 0 {~~} at {~~}\theta = \pi)$,i.e,
this transformation is not a diffeomorphism. Similarly 
the transformation $(\theta, \phi) \rightarrow (U,V)$
is not a diffeomorphism.

We now consider the Robertson-Walker cosmological model.
The space-time metric in terms of comoving isotropic
observers is:

\be
ds^2 = -d{\tau}^2 + a^{2}(\tau)[d{\psi}^2
+ {{\sin^2}{\psi}}(d{\theta}^2 + {\sin^2{\theta}}d{\phi}^2)
\ee
 
here $0 \le \psi \le \pi$,  
$0 \le \theta \le \pi$ and $0 \le \phi \le 2\pi$. The
constant-time
spatial three surfaces ${\sum}_{\tau}$ are compact (topologically $S^3$) 
and there is no four-dimensional spatial geometry available
to embed ${\sum}_{\tau}$. 
The spatial
metric is singular 
along the closed line elements $\theta = 0$,
$\theta = \pi$ including the two point-poles $\psi = 0, \pi$.

The discussions in section IV [equ.(20)] and in this section
show that metric singularities cannot be removed
by diffeomorphically equivalent coordinate transformations.
Thus the black hole and the cosmological
metric singularities are unavoidable aspects of nature.

\section{Appendix: F}

In this section we  consider the electrostatic potential of a
polarized dielectric system. The electrostatic potential
of a polarized dielectric system is given by,
 
\be
V = {1\over{4\pi \epsilon_0}}{\int{{\vec{P}.\hat{R}}\over{R^2}}dv}
\ee

here $\vec{P}$ is the polarization vector of the dielectric material
and $\vec{R}$ is the vector joining the infinitesemal volume element
$dv$ carring a dipole moment $\vec{P}dv$ to the point of observation.
It's magnitude is given by eqns.(46). 
We can reduce equ.(55) to a simpler form consisting two terms:
one from a bound surface charge density $\sigma_b$ 
and another from a bound volume charge density $\rho_b$,

\be
V = {1\over{4\pi \epsilon_0}}{\int_{surface}{{\vec{P}.\vec{da}}\over{R}}}
- {1\over{4\pi \epsilon_0}}{\int_{volume}{{{\vec{\nabla}}.\vec{P}}dv\over{R}}}
\ee  

Here $\sigma_b = \vec{P}.\hat{n}$, $\hat{n}$ is the normal to
the surface of the material and $\rho_b = -{\vec{\nabla}.\vec{P}}$.
 
The total volume charge density in presence of 
a polarized dielectric medium is given by:

\be
\rho = \rho_b + \rho_f + \rho_{sb}
\ee

where we include $\sigma_b =  \vec{P}.\hat{n}$ 
in the free volume charge density
as $\rho_{sb}$
through the introduction of a proper delta function. For example
in the case of a dielectric sphere we have,

\be
{\rho_{sb}{(r,\theta,\phi)}} = {\sigma_b{(\theta,\phi)}} {\delta(r - r_s)}
\ee

Here $r, r_s$ are radial distance (not vectors) and the delta function
have dimension inverse of length.

We then have:

\be
{\epsilon_0}{\vec{\nabla}.\vec{E}} = \rho = \rho_b + {\rho'}_f
\ee

where ${\rho'}_f = \rho_f + \rho_{sb}$ and the divergence of
the eletric
displacement vector $\vec{D}$ is given by,

\be
{\vec{\nabla}.\vec{D}} = {\rho'}_f
\ee

Everywhere apart from the surface of the dielectric we have
${\rho'}_f = \rho_f$ and the above equation (51) 
maches with the conventional exprssion for the divergence
of  $\vec{D}$:

\be
{\vec{\nabla}.\vec{D}} = \rho_f.
\ee
 
The effect of $\rho_{sb}$ should be taken into the boundary
conditions for $\vec{D}$ . This will also have consequences
to obtain the enegy density of a given electrostatic configuration
in presence of dielectric mediums [23] as,

\be
{\vec{\nabla}.\vec{D}} - {\vec{\nabla}.\vec{D_0}} = \rho_{sb}
\ee

We conclude this section with a few comments regarding
the electrostatic field energy in presence of dielectrics.

The eletrostic field energy in presence of dielectric
mediums can approximatly be considered to consist of
three parts [24]:

\be 
W_{tot} = {1\over 2}{{\int D.E}d\tau} = {\epsilon \over 2}{{\int E.E}d\tau}
= W_{free} + W{spring} +  W_{bound}
\ee

here $\epsilon = \epsilon_0{(1 + \chi_e)}$. We briefly explain the
three terms considering the realistic case of a dielectric
filled charged parallel-plate capacitor:

i) $W_{free}$ is the energy to charge the capacitor to produce
the configuration with a given electric field. We can regain
this energy if we discharge the capacitor by connecting the 
two plates through a conductor.

ii) $W{spring}$ is the energy required 
to increase the atomic/molecular dipole moments
or to polarize the atoms/molecules
depending on, respectivly, whether the atoms/molecules have permanent
dipole moments or not. This energy will be regained as heat
when we discharge the capacitor.

iii) $W_{bound}$ is the enrgy required to polarize the dielectric
as a whole. 
The dipole-dipole interaction energy for two dipoles with
dipole moments $\vec{p_1}$ and $\vec{p_2}$ and separated by
$\vec{r}$ is:

\be
U = {1\over{4\pi\epsilon_0}}{1\over{r^3}}
[{\vec{p_1}}.{\vec{p_2}} - 3({\vec{p_1}}.{\hat{r}})({\vec{p_2}}.{\hat{r}})]
\ee
 
$U$ is minimum when the dipoles are antiparallel and
maximum when the dipoles are parallel. 
Consequently for any statistically
infinitesemal volume of the dielectric (i.e, volume elements
which are very small compared to the dimension of system but
large enough to contain sufficient number of atoms/molecules
so that microscopic fluctuations can be approximately averaged to
zero) the orientation of the atomic/molecular dipoles will be as  
isotropic as possible. To polarize the dielectric we have to
orient the atomic/molecular dipoles in 
near-parallel configuration in a given direction
and supply energy to increase the electrostatic energy of the dielectric.
This energy, $W_{bound}$, will be regained as heat if we discharge
the capacitor.

\section{{Supplement:I}}

We will now study the behaviour of of the 
spectrum of covariant Klien-Gordon
equation in the near
horizon limit.

We will first consider the spectrum of the covariant Klien-Gordon
equation  
in the (3 + 1)-dimensional constant curvature black
hole background which contains a one dimensional fixed point set
of the time-like Killing vector field. This black hole space-time
was obtained by Prof. M. Bannados, Prof. R. B. Mann and 
Prof. J. D. E. Creighton through the identification of points
along the orbits of a discrete subgroup of the isometry group of
the anti-de Sitter apce-time. They used a static coordinate
system where
the constant-time foliations become degenarate along a particular
direction apart from the black hole event horizon giving a one-dimensional
fixed point set of the time-like Killing vector. The metric in
the Schwarzschild like coordinates is given by,

\be
ds^{2} = {l^{4}f^{2}(r)\over r^{2}_h}
[d{\theta}^{2} - \sin^{2}{\theta}(dt/l)^{2}]
+ {dr^{2}\over f^{2}(r)} + r^{2}d{\phi}^{2}
\ee
where ${f(r) = ({{{r^{2}} - {r^{2}_h}}\over{l^{2}}})^{1\over2}}$.
These coordinates are valid outside the horizon $(r>r_{h})$
for $0\le{\theta}\le{\pi}$
and $0\le{\phi}<2{\pi}$.  
It is clear that the constant-time foliation becomes degenerate
along the direction  ${\theta} = 0$ and ${\theta} = {\pi}$
giving to a one-dimensional fixed point set of the time-like
Killing vector field.

The covariant wave equation of a minimally coupled massive scalar
field is given by,

\be
{1\over\sqrt{-g}}
{\partial_{\mu}}({\sqrt{-g}}{g^{\mu\nu}}{\partial_{\nu}{\psi}})
-m^{2}{\psi} = 0.
\ee

The solution of the angular equation is given by,

\be
{P^{\mu}_{\nu}(x)} = 
{1\over{{\Gamma}({1 - {\mu}})}}{{({{1+x}\over{1-x}})}^{{\mu}\over2}}
{F(-{\nu} , {{\nu} + 1}; {1-{\mu}}; {{1-x}\over2})}
+ c.c
\ee

where $ x = \cos{\theta}$,
$\mu = iEl$ and $(\nu + \mu) \neq$  an integer. Here
${F(-{\nu} , {{\nu} + 1}; {1-{\mu}}; {{1-x}\over2})}$ is the
hypergeometric function. This solution is $C^1$ throughout the
angular range $0\leq{\theta}\leq{\pi}$. Consequently the energy
spectrum is continuous with divergent density of states.

We will now illustrate that the divergent density of states is a characteristic
feature of the fixed point set of the time-like Killing vector field
indicating the breakdown of the canonical formalism of the
conventional quantum mechanics.

We will illustrate this feature
in the context of the Schwarzschild black hole which contains a
two dimensional fixed point set (the event horizon) of the time-like
Killing vector field. Since the Hawking radiation through which
the non-unitary black hole evaporation takes place originates mostly
from the near horizon region we will consider the behaviour of the
spectrum of the covariant K-G equation
in the near horizon region.
Since the space-time foliation is static we will consider the
stationary states. We will consider the radial solution
of the covariant K-G equation.
The radial
solution can be obtained through the WKB approximation. 
However we can not use the semi-classical quantization
condition.
For the Schwarzschild black hole the constant-time foliations
become degenarate at the black hole event horizon and it is
not possible to impose any consistent boundary condition
on the horizon.
To obtain the degenaracy of the energy eigenstates we will
now consider the radial part of
a covariant generalized probability current density
equation for the low energy eigenstates.
For a general state composed out of superposition
of different energy-eigenstates we consider
the cross-term
taken between states with
neighbouring energy eigenvalues $E$ and $E + \delta E$. 
This gives us the following relation between $E$ and $E + \delta E$:

\be
\partial_{\tau}[Re({{\psi_1}^*}{\psi_2})] =
{1\over m}{\partial_{s}{[Re({{\psi_1}^*}{\partial_{s}{\psi_2}})]}}  
\ee

where the derivatives are taken w.r.t proper time and proper
distance. This expression is similar to the probability
current density equation of unitary quantum mechanics
in presence of damping potentials. This equation is used
because of observed decay, using conventional
quantum mechanics, of the density of states with the proper
distance from the black hole event horizon.

We will obtain the density of states of the energy eigenfunctions
by considering the consistency of the integrated form of the generalized
probability current density equation term by term in an
infinitesemally thin spherical
shell surrounding the black hole with radius $2M + h$  
and $2M + 2h$ where $h~ << ~2M$.
We obtain the following expression for the density of states:

\be
{1\over{\delta E}} = {mK \over {E^2 s}}
\ee

where $s$ is the proper distance between the horizon
and the spherical shell. As $s \rightarrow 0$ the density
of states diverges and the generalized covariant current density
equation becomes consistent. This divergent density of
states is a property of the fixed point set of the time-like Killing
vector field and this density of states gives vanishing
internal energy and entropy for the spectrum of the covariant
K-G equation. 

The continuous energy spectrum is also obtained when one
considers the behaviour of matter fields in the Taub-NUT 
space-time which contains a zero dimensional fixed point 
(in the Euclidean sector) of the time like Killing vector field.
In this case the angular solution (in the Lorentzian sector)
satisfies the minimum regularity condition,i.e, the angular
part of the generalized probability current density integrated
over $S^2$ is finite. This angular solution is similar
to the spin-spherical harmonics.  

The non-unitarity (decay
of density of spectrum with distance
away from the horizon) discussed above is a characteristic aspect of
both the black hole event horizon and the cosmological event
horizon.

We now make some comments regarding relativistic
quantum mechanics similar to App:D. For relativistically
covariant normalization of the quantum mechanical
wave function (or each component of the wave fuction
for spinors) of we have,

\be
{\int{{\psi^*}{\psi}d^4{x}}} = 1
\ee

This indicates that in the reduced units $(c, {h \over {2\pi}} = 1)$
the dimension of $\psi$ is inverse length squared
$[l^{-2}]$. While the lagrangian leading to the 
Klien-Gordon equation is given by:

\be  
L = {{1\over 2}{{\partial \psi^*}\over {\partial x_{\mu}}}
{{\partial \psi}\over {\partial x^{\mu}}}} - {{m^2 \over 2}{\psi^*}^2
{\psi}^2}
\ee

The action determining the space-time evolution
of $\psi$ is dimensionless
in the reduced units. This gives the dimension
of $\psi$ to be $[l^{-1}]$ in contradiction to
that $([l^{-2}])$ obtained from the normalization
condition.

In passing we note that 
to have a consistent time orientation for any
space-time manifold, which through the principle
of equivalence is form of existence of matter fields,
particles should follow a particular family 
of reparameterization invariant
curves in the space-time manifold [5].

\section{{SupplementII: Comments On Classical Electrodynamics}}

In this section we will make a few comments regarding
the basic laws of Classical Electrodynamics.

We first consider the electrostatic potential
of an extended but localized charged distribution.
We will consider points outside the source. Let
$\vec{r'}(r',{\theta}',{\phi}')$ be the position vector for an infinitesemal
volume element $dv'$ within the source which makes an angle
${\theta}'$ with the positive $Z$ polar axis
and an azimuthal angle ${\phi}'$ w.r.t the positive $X$ axis. 
Let $\vec{r}(r,\theta,\phi)$
be the position vector of the point of observation (P) making
an angle $\theta$ with the polar axis
and an angle $\phi$ with the positive $X$ axis. The 
magnitude of the position vector
$\vec{R}$ between $dv'$ and P is then given by:

\be
R^2 = r^2 + {r'}^2 - 2r{r'}cos(\gamma) 
\ee

where,

\be
\cos(\gamma) = {\cos{\theta}'}{\cos\theta} +
{\sin{\theta}'}{\sin\theta}\cos(\phi - {\phi}')
\ee

The electrostatic potential at P is given by,

\be
V(P) = {1\over {4\pi \epsilon_0}}{\int{{\rho(r',{\theta}',{\phi}'){r'^2}
{\sin\theta d{r'} d{\theta}' d{\phi}'}}\over
{[r^2 + {r'}^2 - 2rr'\cos(\gamma)]^{1/2}}}}
\ee

For points outside the source the denominator can be binomially 
expanded in terms of Legendre
polynomials of $\cos(\gamma)$ . Using the addition theorem for Legendre polynomials:

\be
P_l{(\cos{\gamma})} = P_{l}(cos{\theta})P_{l}(cos{\theta}') + 
2\sum{{(l - m)!}\over{(l + m)!}}{{P_l}^m(\cos{\theta})}{{P_l}^m(\cos{\theta}')}
{\cos[m(\phi - {\phi}')]}
\ee

the denominator can be expanded in terms of products of Legendre polynomials of 
$\cos{\theta}$
and $\cos{\theta}'$ [23]. For a spherically symmetric charge distribution we have only the monopole
term $Q \over{4\pi \epsilon_0 r}$
from this expansion for points with $r > r'$.
However if we consider points near the South pole, the convensional binomial
expansion [24] is not valid in general for $({{r'}\over{r}}) \geq {\sqrt{2} - 1}$ 
although the series expressed in
terms of the Legendre polynomials converge for these points
(This aspect are also partly discussed
in [28]). The convergence of the series
does not justify the binomial expansion.
To illustrate let us consider the potential of a point
charge, $q$, situated at $z = a$. The potential expanded in terms
of Legendre polynomials for $r > a$ is given by:

\be
V(\vec r) = {q\over {4\pi \epsilon_0}}
\sum{{{a^l}\over{r^{l + 1}}}{P_{l}(cos{\theta})}}
\ee

For points on the negative $Z$-axis and infinitesimally
close to $z = -a$ we have:

\be
V(\vec r) = {q\over {4\pi \epsilon_0 a}}
[1 - f(\delta)]
\ee

here $\delta = {\epsilon\over a}$,
$r = a + \epsilon$ and we have kept only
terms linear in $\delta$. Whereas Coulomb's
law give us the following expression for
electrostatic potential at $z = -(a + \epsilon)$:

\be
V(\vec r) = {q\over {8\pi \epsilon_0 a}}
[1 - {(\delta)\over 2}]
\ee
  
here also we have kept terms linear in 
$\delta = {\epsilon\over a}$. Clearly the potential obtained at $z = -a$
from expansion in terms of Legendre polynomials
is not in accordance with Coulomb's law.
Whereas the potential obtained from eqn.(71) is given
by,

\be
V(\vec r) = {q\over {8\pi \epsilon_0 a}}
[1 - {{(\delta)^2}\over 2}]
\ee

We should also note that Legendre polynomials
are either symmetric or antisymmetric around
$\theta = {{\pi}\over{2}}$ and for a charge distribution
on an arbitrary shaped conductor which can not be expressed as a sum
of a symmetric and an antisymmetric part we
can not expand the corresponding potential
in terms of Legendre polynomials.

In general, for any point outside the source we can use two consequitive binomial expansion:
the first one is factoring out 
$(r^2 + {r''}^2)$ in the denomenator
of $V(P)$ and performing a binomial series expansion in
terms of $2{{r r'}\over{(r^2 + {r'}^2)}}cos(\gamma)$.
The second one is in terms of ${{r'}^2}\over{r^2}$
as is evident from the following expresion:

\be
V(P) = {1\over {4\pi \epsilon_0}}{\int{{\rho(r'){r'^2}
{\sin\theta d{r'} d{\theta}' d{\phi}'}}\over
{{(r^2 + {r'}^2)}^{1/2}
[1 - 2{{r r'}\over{(r^2 + {r'}^2)}}\cos(\gamma)]^{1/2}}}}
\ee

In this expression any power of $(r^2 + {r'}^2)$ can be binomially expanded in terms of 
${{r'}^2 \over {r^2}}$
for all $r > r'$. This expansion gives the usual results for the monopole potential term and
the dipole potential term (which vanishes for a spherically symmetric source). However for a
spherically symmetric charge distribution we get a non-trivial screening term. To the order
of $1\over{r^5}$ this term is given by:

\be
V({1\over{r^5}}) = {1\over {4\pi \epsilon_0}} 
\int{\left\{-{{3(r')^4}\over{8 r^5}} + {{7(r')^5}\over{32 r^5}}\left[
{2\over 5}({P_2{(\cos\theta)}})^2 + {1\over 4}\right]\right\}}
{{r'}^2{\sin{\theta}'}{\rho(r')}d{r'} d{\theta}' d{\phi}'} 
\ee

The $\theta$-dependent term of the corresponding electric field
vanishes at $\theta = {\pi \over 2}$ and the angular component is
directed towards the equetorial plane. These aspects are valid for
all the $\theta$-dependent terms for a spherical charge distribution.

The potential of a charged spherical shell for points infinitesimal
close to the surface of the shell is:

\be
V(P) = {{{{r_s}^2}\sigma}\over {4\pi \epsilon_0 r}}\int{
{\sin{\theta}' d{\theta}' d{\phi}'}\over
{[2(1 - \cos{\gamma}) + {{2\epsilon}\over{r_s}}(1 - \cos{\gamma})]^{1/2}}}
\ee

We know from the electrostatic properties of conductors that the 
the out-side electric field on the surface of the
conductor is,

\be
\vec{E} = {{\sigma}\over{\epsilon_0}}\hat{r} 
\ee

This leads us to the following expression:

\be
\int{{\sin{\theta}'d{\theta}' d{\phi}'}\over 
{[2(1 - \cos{\gamma})]^{1/2}}} = 8 \pi 
\ee

We will also have two quadratures with value zero.

However the expansion corresponding to equation (71) differs from the
conventional expansion in terms of Legendre polynomials and the
uniqueness theorems of Electrostatics lead us to conclude that
one of these expansions are valid.

Secondly the presence of the screening term indicate that the integral
version of Gauss's divergence law in Electrostatics is in general
not valid. The non-validness of the integral version of Gauss's
divergence law in Electrostatics is also evident for the electric field
of an infinite charged cone [23] .
We can also construct some charge distribution
with a conductor, e.g,
two finite radial cone joined through a spherical knot 
with center at the origin. For such charge distributions
the potential near the conductor cannot be expressed in
terms of Legendre polynomials and it is expected that
that the integral version of Gauss's law for electric field will 
not be valid.
Non-validness of the integral version of Gauss's divergence
theorem may be associated with the discontinuous nature
of the source.

We will now rederive Maxwell's equations without
using the integral version of Gauss's divergence law
for fields with localized sources.

We first consider Gauss's law for the electrostatics.
For a point charge at the origin the surface integral of the static 
electric field, determined by an inverse square law,
over a closed sphere centered at the origin is $Q/{\epsilon_0}$. 
This is because the origin of
the coordinate system coincides
with the point charge and the screening terms
in eqn. (72) are 
vanishing. 
Consequently the charge density in spherical polar coordinates
is given by: $\rho(r) = {Q\over {4\pi r^2}}{\delta(r)}$. 
Here $\delta(x)$
is the Dirac-delta function and we have the following
useful relation:

\be
[{{{\vec{\nabla}}_{\vec{r}}}}.{({{\hat{r}}\over{r^2}})}]
= 4\pi{\delta^{3}(\vec{r})}
\ee

We now consider the divergence of $\vec{E}$ inside
source. To calculate the divergence at a point 
$\vec{r}$ within the source we break the source into
two parts: one is an infinitesimal spherical volume element
of radius ${r'}_s$ centered at $\vec{r}$ and the other is
the rest of the source. The electric field is sum of
two parts: one due to the infinitesimal volume element
($\Delta v$), ${\vec{E}}_{\Delta v}$, and the other due to the   
rest of the source,  ${\vec{E}}_{rest}$. 
To calculate the divergence of ${\vec{E}}_{\Delta v}$
we can use a spherical coordinate system centered at 
$\vec{r}$. The position vector
is given by $\vec{r''} - \vec{r}$.
The boundary of ($\Delta v$) is given
by $|\vec{r''} - \vec{r}| = {r'}_s$. 
We also break the charge density into two
parts:

\be
\rho({\vec{r'}}) = {\rho_{\Delta v}}({\vec{r''}}) 
+ {\rho_{rest}}({\vec{r'}})
\ee

${\rho_{\Delta v}}({\vec{r''}})$ is non-zero
$[= \rho({\vec{r'}})]$ for points within ${\Delta v}$
$[|\vec{r''} - \vec{r}| < {r'}_s]$
while ${\rho_{rest}}({\vec{r'}})$ is non-zero
$[= \rho({\vec{r'}})]$ for points not within
${\Delta v}$ $[|\vec{r'} - \vec{r}| \geq {r'}_s]$.

We now
have the following expression for the divergence
of ${\vec{E}}_{\Delta v}$:

\ba
{4 \pi \epsilon_0}{\vec{\nabla}}.{{\vec{E}}_{\Delta v}} 
& = & \int_{\Delta v}{{\rho_{\Delta v}}({\vec{r''}})}
[{{{\vec{\nabla}}_{\vec{r}}}}.{({{\hat{R}}\over{R^2}})}]
{(|\vec{r''} - \vec{r}|)^2}{d|\vec{r''} - \vec{r}|}
{\sin{\theta''}}{{d\theta''}{d\phi''}}\nonumber\\
& = & -\int_{\Delta v}{{\rho_{\Delta v}}({\vec{r''}})}
[{{{\vec{\nabla}}_{(\vec{r''} - \vec{r})}}}.{({{\hat{R}}\over{R^2}})}]
{(|\vec{r''} - \vec{r}|)^2}{d|\vec{r''} - \vec{r}|}
{\sin{\theta''}}{{d\theta''}{d\phi''}}\nonumber\\
& = & \int_{\Delta v}{{\rho_{\Delta v}}({\vec{r''}})}
[{{{\vec{\nabla}}_{(\vec{r''} - \vec{r})}}}.{(-{{\hat{R}}\over{R^2}})}]
{(|\vec{r''} - \vec{r}|)^2}{d|\vec{r''} - \vec{r}|}
{\sin{\theta''}}{{d\theta''}{d\phi''}}
\ea

Here $\vec{R} = (\vec{r} - \vec{r''})$ and to obtain
the first expression we have used the fact that
${\rho}$ do not depend on the
unprimed coordinates.
As we had discussed in the last section
${{{\vec{\nabla}}_{(\vec{r''} - \vec{r})}}}.{(-{{\hat{R}}\over{R^2}})}
= 4\pi{\delta^{3}(\vec{r''} - \vec{r})}$. It is now
straight forward to show that the divergence of $\vec{E_{rest}}$
vanishes for points within ${\Delta v}$.
Thus we have
Poisson's equation for points inside an extended
source:

\be
{\vec{\nabla}}.{\vec{E}(\vec{r})} = {{\rho(\vec{r})}\over{\epsilon_0}}
\ee

For a volume charge density $\rho(\vec{r})$ the
charge density should vanish at the surface
of the source. Otherwise we will have a non-trivial
surface charge density.
For non-trivial surface and line charge densitis the divergence
of $\vec{E}$ can be found following the above
procedure and the results are same as replacing
the source through proper delta functions.

As usual the curl of $\vec{E}$ is zero.

The no-work law can be regained for the electrstatic field
of an extended charge distribution if we generalize
the derivation properly. The line-integral of $\vec{E}$
over a closed contour for each element
of the source with the source-coordinate remaining
fixed is given by:

\be
{\int_{contour}{d\vec{E}}.{\vec{dl}}} =  
{\int_{contour}{{\partial{[dV(R)]}}\over{\partial{R}}}}{dR} = 0
\ee

and the total work done for the whole source is obviously zero.

We now consider the divergence and curl of the
magnetostatic field: $\vec{B}$. The Biot-Savart
law for the general case of a volume current density
$\vec{J}$ is given by:

\be
{\vec{B}} = {{\mu_0}\over{4\pi}}{\int{{\vec{J}{(r',\theta , \phi)} 
\times \hat{R}}
\over {R^2}}{{r'}^2 {\sin{\theta}}dr' d\theta d\phi}}
\ee

where $R$ is given by eqn.(63).

It can easily be shown that ${\vec{\nabla}.{\vec{B}}} = 0$
following the conventional procedure [24].

The curl of $\vec{B}$ is given by,

\be
{\vec{\nabla} \times {\vec{B}}} = {\mu_0}{{\vec{J}}{(r',\theta,\phi)}}
- {\mu_0 \over {4\pi}}\int{{(\vec{J}.\vec{\nabla}){\hat{R}\over{R^2}}}dv}
\ee

Here the integration is over the source volume.
The first term 
arises from the source volume integrand
(apart from a multiplicative factor):
${{\vec{J}}{{\vec{\nabla}}.({\hat{R}}/R^2)}}$.
Following the same procedure as to obtain
the ${{\vec{\nabla}}.{\vec{E}}}$ law
we obtain the first term
in the above equation.

We now
consider the second term. We have, for the $x$-component,

\be
\int{{(\vec{J}.\vec{\nabla}){(x - x') \over{R^2}}}dv} =
\int{{\vec{\nabla}}.[{{(x - x') \over{R^3}}{\vec{J}}}]dv} - 
\int{{{(x - x') \over{R^3}}({\vec{\nabla}}.{\vec{J}})}dv}
\ee

The second term in the r.h.s of
the above equation vanishes as $\vec{J}$ do not depend
on the unprimed variables.
The first term is given
by,

\be
\int{{\vec{\nabla}}.[{{(x - x') \over{R^3}}{\vec{J}}}]dv}
= -\int{{\vec{\nabla'}}.[{{(x - x') \over{R^3}}{\vec{J}}}]dv}
\ee

This is possible because for steady currents
${\vec{\nabla}'}.{\vec{J}} = 0$.
The integration gives terms dependent on $\vec{J}$
on the boundary of the source. 
This is apparent if we use the Cartezian coordinate
system.
As discussed in the
context of ${\vec{\nabla}}.{\vec{E}}$ law, $\vec{J}$
should 
vanish on the boundary of the source otherwise we will have
a non-trivial surface current density.
It can easilly be shown that 
for an arbitrary curved space with line element:
$ds^2 = {h_1}^2{({x_1}, {x_2}, {x_3})}(d{x_1})^2
+ {h_2}^2{({x_1}, {x_2}, {x_3})}(d{x_2})^2
+ {h_3}^2{({x_1}, {x_2}, {x_3})}(d{x_3})^2$
the above integral vanishes for the same
boundary condition as above on $\vec{J}$.

We will now consider surface current density.
Let us first consider a surface with an wedge, i.e, two
surfaces joined along a curve making
an angle which can, in general, vary along the
curve. If a surface current density originates
at one surface the current can not propagate
to the second surface as the 
only way that the current can have vanishing
perpendicular components to both the surfaces
is to propagate along the wedge
(geometrically for two surfaces making
an wedge the only way that the ideal
surface current density can remain tangential
to both the surfaces at the wedge is to
flow along the wedge).
The above arguments
demonstrates that we can not have surface currents
out of point charges flowing along an arbitrarilly shaped surface
as an arbitrary surface can be expressed as a collection
of infinitesimal flat srfaces with non-parallel normals.
Similar arguments as above demonstrates that
we can not have currents in an arbitrary curve
out of motion of point charge carriers.

We can have surface currents for an infinite plane
surface and provided the surface current density
is not divergent the boundary term in
eqn.(84) vanishes as the factor $(x - x')\over{R^3}$ 
vanishes for $x' \rightarrow {\pm \infty}$. We can
also have steady currents only
out of axial rotation of an azimuthally symmetric
surface charge distribution. In this case, expressed
in terms of cylindrical polar coordinates or
spherical polar coordinates the boundary term in
equ.(84) vanishes out of uniqueness. Similar arguments
show that for line-current densities, which can be
either an infinite straight line current or azimuthally
symmetric rotation of a charged ring, the boundary
terms in eqn.(84) vanishes.

Thus we have,

\be
{\vec{\nabla} \times {\vec{B}}} = {\mu_0}{{\vec{J}}{(r',\theta,\phi)}}
\ee

For ideal surface and line current densities the results will
same as replacing ${\vec{J}({\vec{r}})}$ 
by suitable delta functions meausured on the the source
provided the sources satisfy the required regularity
conditions as discussed above.

The well-known integral law for a physical line current
density 
$\int_{contour}{\vec{B}.\vec{dl}} = {\mu_0}I$,
where the contour is a closed circle concentric
with the source and lies on a plane perpendicular
to the physical line-source, can be easily derived
following the procedure used to establish
the no work law for the electrostatic field
although for a physical line-source $\vec{B}$
will have a small non-vanishing 
radial component
on the plane of the circle.
To illustrate a physical line current density
can be regarded as a collection of parallel infinitesimal
line-current density for each of which ampere's circuital
law is valid along their axeses.

Faradey's law, with the following convention (Section:7.1.3 [24])  

for motional emf or induced electric field the
direction of the current or the electric field
along a closed loop and the orientation of
the enclosed surface
giving the magnetic flux are related by the right-hand
thumb rule,

together with the above discussions
and the current density equation (differential
version of the electric charge conservation law)
reproduces Maxwell's
laws of Classical Electrodynamics. After that we have
reestablished 
Maxwell's equations
we should note that
the finite volume
of the elementary charge carriers indicates
that ideal line/surface charge densities
cannot exist in nature unless the charge carriers
can move with velocity $c$ in one or two directions.  
Also to have ideal line/surface current densities 
the charge carriers should move with speed
greater than $c$. Point charges can not
exist in nature and the electromagnetic self-energies of the
elementary particles are not infinite. Also considered as a classical
model the finite
volumes of the elementary charges give rise to
screenining terms as discussed in the context
of the electrostatic potentials of extended
charged systems.

We conclude this article with a few comments on magnetic monopoles.
It is straight forward to show that under a general
electromagnetic duality
transformation, eqn.(6.151, 6.152) [23],

${\vec E} = {\vec{E'}}{\cos{\xi}} + {\vec{H'}}{\sin{\xi}}
~~~~~~~~{\vec D} = {\vec{D'}}{\cos{\xi}} + {\vec{B'}}{\sin{\xi}}$

${\vec H} = {-\vec{E'}}{\sin{\xi}} + {\vec{H'}}{\cos{\xi}}
~~~~~{\vec D} = {-\vec{D'}}{\sin{\xi}} + {\vec{B'}}{\cos{\xi}}$
 
and 

${{\rho}_e} = {{\rho}'_e}{\cos{\xi}} + {{\rho}'_m}{\sin{\xi}}
~~~~~~~~{\vec{J_e}} = {\vec{J'_e}}{\cos{\xi}} + {\vec{J'_m}}{\sin{\xi}}$

${{\rho}_m} = {-{\rho}'_e}{\sin{\xi}} + {{\rho}'_m}{\cos{\xi}}
~~~~~{\vec{J_m}} = {-\vec{J'_e}}{\sin{\xi}} + {\vec{J'_m}}{\cos{\xi}}$

the differential version of the magnetic charge
conservation law doe's not remain time-reversal symmetric
due to the pseudoscalar and pseudovector nature
of magnetic monopole charge and magnetic current
density vector respectivly, i.e, if 
${\vec{\nabla}}.{\vec{J'_m}} = - {{\partial{\rho}'_m}
\over{\partial t}}$ is time-reversal
symmetric then 
${\vec{\nabla}}.{\vec{J_m}} = - {{\partial{\rho}_m}
\over{\partial t}}$ no-longer remains time-reversal
symmetric although the differential version
of the electric charge conservation law remains time-reversal
symmetric under electromagnetic duality transformation.

\section{{SupplementIII: Comments on Hydrodynamics}}

In this article we will review the laws of fluid
dynamics. Our discussions will be based on mainly
that of chapter 40, 41 of The Feynman Lectures on
Physics, Vol.2 [25]. 

The dynamics of dry water is governed by eqn.(40.6) [25]:

\be
{\partial{\vec{v}}\over{\partial t}} 
+ ({\vec{v}}.{\vec{\nabla}}){\vec v} =
-{{{\vec{\nabla}}p}\over{\rho}} - {{\vec{\nabla}}{\phi}}
\ee

or using a vector analysis identity to the second
term of the above equation:

\be
{\partial{\vec{v}}\over{\partial t}} 
+ {({\vec{\nabla} \times {\vec{v}}}) \times {\vec{v}}}
= -{{{\vec{\nabla}}p}\over{\rho}} - {{\vec{\nabla}}{\phi}}
\ee

where $\vec{v}$ is the velocity of an fluid element
for which such laws can be applicable, $p$ is the 
fluid pressure and $\phi$ is the potential per
unit mass for any potential force present.
We can derive some important laws from eqn.(86).
The first one is the equation for vorticity
$(\Omega = {\vec{\nabla} \times {\vec{v}}})$
and is obtained by taking curl of eqn.(87):

\be
{\partial{\vec{\Omega}}\over{\partial t}} 
+ {{\vec{\nabla}} \times ({\vec{\Omega}} \times {\vec{v}})}
= 0
\ee

The second one is Bernoulli's theorems (40.12) and (40.14) [26]:

\be
{\vec{v}}.{\vec{\nabla}}({p\over \rho} + \phi + {1\over 2}{v^2}) = 0
\ee

i.e,

\be
{p\over \rho} + \phi + {1\over 2}{v^2} = 
const {~} (along {~} streamlines)
\ee

valid for
steady flow and

\be
{p\over \rho} + \phi + {1\over 2}{v^2} = 
const {~} (everywhere)
\ee

valid for steady and irrotational flow.

However in all these equations the variation of
the fluid density, $\rho$, is not considered while
in deriving eqn.(40.17) [25] the variation of
fluid density is not properly taken into account.
The consideration for 
variation of the density of a nearly-incompressible
fluid may become important through the facts that
when unconstrained the shape of a fluid can be
changed almost freely and sparsed away and through the facts that
layers of fluids can be very easilly spreaded or dettached
away although these properties vary from fluid to
fluid.
These features together with the
local version of the conservation of mass law
(assuming that there is no local source or sink
in the region of interest):

\be
{\vec{\nabla}}.(\rho \vec{v}) = 0
\ee

indicate that we should consider the 
possibility for variation of
$\rho$ properly as we will illustrate later
that some ideal models can cause a finite
variation
of $\rho$ and in reallity the description
of the motion should be changed. 
While the divergence of $\vec{v}$ 
may become important
in cases like Couette flow where the centrifugal
forces imposes a finite and may even be large
divergence of $\vec{v}$.

We can derive a proper version of eqn.(86) by applying
Newtons second law to the fluid momentum per unit volume
and we have:

\be
{\partial{(\rho\vec{v})}\over{\partial t}} 
+ [{\vec{v}}.{\vec{\nabla}}]{(\rho\vec v)} =
-{{\vec{\nabla}}p} - {{\vec{\nabla}}{(\rho\phi)}}
\ee

This equation is in general a non-linear coupled
[through eqn.(92)] partial differential
equation for $\vec{v}$.

Bernoulli's theorems for
fluid dynamics can only be established when
$\rho$ is constant :

\be
{\vec{v}}.{\vec{\nabla}}[p + (\phi \rho) + 
{1\over 2}{(\rho v^2)}] = 0
\ee

i.e,

\be
p + (\phi \rho) + 
{1\over 2}{(\rho v^2)} = 
const {~} (along {~} streamlines)
\ee

valid for
steady flow and

\be
p + (\phi \rho) + {1\over 2}{(\rho v^2)} = 
const {~} (everywhere)
\ee

valid for steady and irrotational flow.

In general, when $\rho$ is varying,
only the first of the Bernoulli's theorems :

\be
{\vec{v}}.{\vec{\nabla}}[p + (\phi \rho) + 
{1\over 2}{(\rho v^2)}] = 0
\ee

remains to be valid provided 
$({\vec{v}}.{{\vec{\nabla}}\rho})$
is vanishing or is approximately valid if
$|({\vec{v}}.{{\vec{\nabla}}\rho}){\vec{v}}|$ is
negligible compared
to the other terms in eqn.(93). 
To illustrate the significance of these
comments, let us consider the ideal model to
calculate the efflux-coefficient, fig. 40-7 [25].
After that the contraction of the cross-section of the emerging
jet has stopped we have, from the coservation of
mass law, the following equation for $\rho {v}$
at two vertical points:

\be
{\rho}_1{v_1} = {\rho}_2{v_2}
\ee

In this case pressure is the atmospheric pressure
and remains the same throughout the flow and
thus even for the flow of a nearly-incompressible
fluid $\rho$ can vary as $v$ changes with height. In reality
the flow usually gets sparsed away after a distance
which varies for different flows.

The viscous flow of a fluid
is governed by the following two laws which
are obtained from eqn.(93) and eqn.(41.15),[25]: 

\be
{\partial{(\rho\vec{v})}\over{\partial t}} 
+ [{\vec{v}}.{\vec{\nabla}}]{(\rho\vec v)} =
-{{\vec{\nabla}}p} - {{\vec{\nabla}}{(\rho\phi)}}
+ {\eta}{{{\nabla}^2}\vec{v}} +
({\eta} + {\eta}'){{\vec{\nabla}}({\vec{\nabla}}.{\vec{v}})}
\ee

\be
{\vec{\nabla}}.(\rho \vec{v}) = 0
\ee

supplimented by proper boundary conditions. 
To illustrate the significance of the boundary
conditions we can consider the change of the 
shape of the surface of water in a bucket when
the bucket is given a steady rotational motion
about it's axis. The surface of the water
become paraboloidal when the bucket is rotating.
This shape can not be obtained without
a vertical component of fluid velocity
along the bucket surface
for a finite duration
although the bucket surface only have
an angular velocity. 

In the above equations
$\eta$ is the ``first coefficient of viscosity''
or the ``shear viscosity coefficient'' and
${\eta}'$ is the ``second coefficient of viscosity''.
This equation is extremely significant in the sence
that this equation, not eqn.(41.16) [26], is the 
equation which contains all the terms relevant
to describe the dynamics of  viscous fluids,
both nearly-incompressible and compressible.
For compressible fluids $\rho$ will also
depend on pressure, $p({\vec{r}})$.
We can modify this equation only through
varying the nature of the viscous force.

The equation for vorticity is given by:

\ba
{\partial{\vec{\Omega}}\over{\partial t}} 
+ {{\vec{\nabla}} \times ({\vec{\Omega}} \times {\vec{v}})}
- {{\vec{\Omega}}({\vec{\nabla}}.{\vec{v}})}
+ {{\vec{v}} \times {{\vec{\nabla}}({\vec{\nabla}}.\vec{v})}}
= {{\eta \over \rho}({{\nabla}^2}{\vec{\Omega}})}
{~~~~~~~~~~~~~} & {~} & \nonumber\\
- {{({{\vec{\nabla}}p}) \times ({{\vec{\nabla}}\rho})}
\over {{\rho}^2}}
- {{[{{\vec{\nabla}}{(\rho\phi)}] \times ({{\vec{\nabla}}\rho})}
\over {{\rho}^2}}} 
+ {{{\eta}({{{\nabla}^2}{\vec v}}) \times ({{\vec{\nabla}}\rho})}
\over {{\rho}^2}}
+ {{\eta + {\eta}'}\over {{\rho}^2}}
{{\vec{\nabla}({\vec{\nabla}}.{\vec{v}})} \times ({{\vec{\nabla}}\rho})}
\ea

We can obtain an equation similar to eqn.(41.17) [25]
describing the motion of a viscous fluid past a cylinder
provided we can neglect the terms involving ${{\vec{\nabla}}\rho}$
and it is given by:

\be
{\partial{\vec{\Omega}}\over{\partial t}} 
+ {{\vec{\nabla}} \times ({\vec{\Omega}} \times {\vec{v}})}
- {{\vec{\Omega}}({\vec{\nabla}}.{\vec{v}})}
+ {{\vec{v}} \times {{\vec{\nabla}}({\vec{\nabla}}.\vec{v})}}
= {{\eta \over \rho}({{\nabla}^2}{\vec{\Omega}})}
\ee

Following the procedure
in section 41-3,[26] we can rescale the variables to obtain an equation
which has Reynolds number $(R)$ as the only free
parameter :

\be
{\partial{\vec{\omega}}\over{\partial t'}} 
+ {{\vec{\nabla}'} \times ({\vec{\omega}} \times {\vec{u}})}
- {{\vec{\omega}}({\vec{\nabla}'}.{\vec{u}})}
+ {{\vec{u}} \times {{\vec{\nabla}'}({\vec{\nabla}'}.\vec{u})}}
= {{1 \over R}({{{\nabla}'}^2}{\vec{\omega}})}
\ee

where the prime describe the scaled variables,
$\vec{u}$ is the scaled velocity
and $R$ is given by the usual expression,
$R = {{\eta \over \rho}V D}$.

To conclude in this section we have derived the exact equation
describing fluid dynamics. We considered the motion of both
non-viscous 
and viscous fluids. We proved that in both the
cases there are terms which are neglected in the
conventional theory but may become significant
in some ideal model and 
in reallity the description of motion is changed.
Some of these terms even change the dynamical
laws of
viscous fluid motions by violating the conventional
theory established in term of the Reynold number
and these terms are significant for 
the dynamics of compressible
fluids like air.

\section{{SupplementIV: Comments on Double slit interference}}

In this section we make a few comments regarding double
slit interference experiment following section 13.3 of
[26]. 

In Figure.13F the wave fronts 
that reach at the screen point $P$ simualtaneously
from the 
slits $S_1$ and $S_2$ should have originated from the
two slits at different instants. The wavefront from
$S_2$ should originate at time ${S_{2}A}\over {c}$
earlier than the corresponding interfering wave front
that has originated at $S_{1}$. 
The phase difference is: $\delta = \omega{{S_{2}A}\over {c}}
= {{2\pi}\over {\lambda}}{S_{2}A}$ and we can continue
the corresponding analyses discussed in [27] to obatain the fringe
pattern. Similarly, to reach
a particular screen point $P$ simultaneously the
wavefronts from different parts of an aperture should
have originated from wavefronts at the aperture
at different instants of time differing in phases.
This gives rise to diffarction. These discussions demonstrate
that interference and diffraction are not only
wave phenomena but also associated with  
finite velocity of waves and for electromagnetic waves
in vacuum this velocity is $c$ as is verified in
interference and diffraction experiments. Following the
above arguments the proper expression for
fringe-shift in the Michelson-Morley
experiments [26] is $2{\omega}d{{v^2}\over{c^3}}$,
where $\omega$ is the frequency of light.
We can not follow the procedure followed in [26] to obtain
the path-difference as the velocity of light, as
is assumed, is different in different directions and as discussed
above it is not proper to calculate the phase-difference
between the interfering waves by multiplying the path-dfference
by ${{2\pi}\over {\lambda}}$ directly. 
The fact that we can have the interference fringe
system by allowing one photon to emarge from
the source at each instant indicate that we have to introduce
the corresponding electromagnetic wave description as
fundamental and this description do not depend on
the width of the slits.

These discussions can be continued for electron beam interference
experiments. In this case the wave fronts, the 
position-space wave functions $\psi$, are of the form
$N{e^{i({{\vec k}.{\vec r}} -
{\omega}t})}$,
and
the inerfering states are the states of the electron
at the slits at two different instants of time. Only
for the central maximum the two states 
at the two slits are the same. 

$\psi$ is the complete microscopic description
of the electrons in the electron
-beam interference experiments. It is so as there is no underlying
ensemble representation of the probabilistic interpretation
of $\psi$. 
If we have to assume that, unlike the electromagnetic interference
experiments,
$\psi$ is
vanishing at the slits for slit width less than that
of the diameter of the electrons then the fundamental
reality of $\psi$ is questionable. In other words
$\psi$ gives a microscopic description of the electron
but below a certain length the kinamatic reality of
$\psi$ should be replaced by a proper description.
We can also illustrate this aspect with the following
question:

What is the quantum mechanical description of a Radon
atom in a rigid box when the distance of consecutive
nodes and antinodes of $\psi$ is equal to or 
less than the diameter of the atom?

\section{{Supplement:V(A Few Question)}}

What happens to the entropy increase principle as the
Universe evolve to form the big-crunch singularity?
What happens to the uncertainity relations along the
process of gravitational collapses? 
What is the quantum mechanical description of a Radon
atom in a rigid box when the distance of consecutive
nodes and antinodes of $\psi$ is equal to or 
less than the diameter of the atom?
What is the position-space
wave function of two finite volume massive bosons if we
take contact interaction into account? 
How a photon produce 
electron-positron pair with finite volume concentrate
rest masses?
What are the charges and masses of the electron-positron
pairs forming loops in the vacuum?
How two particles with three-momentums $k_1 , k_2$ $(k_1 \neq k_2)$ 
produced to form a loop
at a space-time point always arrive at another
spacial point simultaneously? 
What is the microscopic
explanation in terms of particle
exchanges of the force in the Casimir effect?
What is the mechanism of the collapse of the momentum-space
wave function of a particle knocking out an elctron from an
atom? What is meant by $|\Psi> = {c_1 (t)}|\Psi_{U^{238}}> +
{c_2 (t)}|\Psi_{Th^{234}}> ~?$ Quantum mechanically 
the region between the rigid walls
(which is equiprobable in classical mechanics) is non-homogeneous
for a particle in a rigid box !
A photon can not reproduce Maxwell's equations
apart from moving with velocity $c$. How can a process
involving only a few photons be described
starting from the Maxwell's equations?
The large scale structure of the Universe is homogeneous.

What is the screen in our brains to view objects, as they are, of sizes
larger than our brains?

\section{{Acknowledgment}}

I am thankful to the Theory Division, S.I.N.P,
Kolkata for their cooperations. I am also thankful to 
Prof. D. J. Griffiths for pointing out a correction regarding
equ.(51). I am also thankful to a member of VECC Housing, Kolkata
for reminding me an important point.

\section{{References}}

[1] J. J. Sakurai;
    Advanced Quantum Mechanics: Addison-Wesley Publishing
    Company.

[2] A. N. Matveev;
    Mechanics and Theory of Relativity: Mir Publishers, Moscow.

[3] R. M. Wald;
    General Relativity: The University of Chicago Press,
    Chicago and London.

[4] C. W. Misner, K. S. Thorne, J. A. Wheeler;
    Gravitation: W. H. Freeman and Company, New York.

[5] A. N. Kolmogorov and S. V. Fomin;
    Functional Analysis (Vol. I): Dover Publications, Inc. New York.

[6] W. Rudin;
    Introduction to Topology and Modern Analysis: 
    Mc Graw - Hill Book Company.

[7] W. Lamb (Jr.); Physics Today 22.

[8] S. Chandrasekhar.

[9] A. K. Raychoudhury.

[10] N. N. Bogoliubov, D. V. Shirkov;
     Introduction To The Theory of Quantized Fields:
     John Wiley and sons.

[11] A. K. Raychoudhury {\it et all \it}.

[12] A. Afriat, F. Sellery;
     The Einstein, Podolosky and Rosen paradox:
     Plenum publishing corporation.

[13] J. J. Sakurai; Modern Quantum Mechanics (Section: 1.4 - 1.6):
     Addison Wesley Publishing Company.

[14] K. Ghosh; gr-qc/0503003.

[15] M. Nakahara; Geometry, Topology And Physics:
     Adam Hilger, Bristol And New York.

[16] T. M. Apostol; Mathematical Analysis: Narosa 
     Publishing House.

[17] S. Weinberg; Gravitation and Cosmology: John Wiley
    and Sons.

[18] M. E. Preskin, D. V. Schroeder; An Introduction to
     Quantum Field Theory: Addison-Wesley Publishing
     Company.

[19] S. W. Hawking, G. F. R. Ellis; The Large Scale Steucture
     of Space-Time: Cambridge University Press.

[20] L. D. Landau and E. M. Lifshitz; The Classical Theory of
     Fields: Butterworth-Heinenann.

[21] J. D. Bjorken, S. D. Drell; Relativistic Quantum Fields:
     Mc Graw-Hill Book Company.

[22] P. Ramond; Field Theory: A Modarn Primer, 
     Addison Wesley Publishing Company

[23] J. D. Jackson; Classical Electrodynamics: Wiley Eastern Limited.

[24] D. J. Griffiths; Introduction To Electrodynamics(1989):
     Prentice-Hall of India.

[25] R. P. Feynman, R. B. Leighton, M. Sands: The Feynman Lectures
     on Physics; Narosa Publishing House.

[26] F. A. Jenkins, H. E. Harvey; Fundamentals of Optics:
     McGraw-Hill Book Company.

[27] R. Longhurst; Geometric and Physical Optics: Orient-Longman.

[28] G. B. Arfken, H. J. Weber; Mathematical Methods For
Physicists: Academic Press, Inc.

\newpage

\section{{Samapta}}

Any one, who had seriously disturbed 
the author  
academically or non-academically
during
the last five years, in particular through
undue slanging and horning out of a dogging
heritage  while the article
was getting prepared, and/or encouraged
to do so is a descendant of Avatar of Dharmaraj.

Reference: Jessy's artificial hand.

\end{document}